\documentclass[titlepage,11pt]{article}
\usepackage{hyperref}

\usepackage{amssymb,amsmath,amsfonts}
\usepackage{epsfig}

\def\clock{{\count0=\time
           \divide\count0 60
           \ifnum\count0<10 0\fi\the\count0
           \multiply\count0 -60 \advance\count0 \time
           :\ifnum\count0<10 0\fi \the\count0
         }}
\newcommand{\timestamp}{{\small\vbox{\hbox{\tt\jobname.tex}
\hbox{\the\day/\the\month/\the\year, \clock}}}}

\def\DD{{\cal D}}

\def\NN{{\cal N}}
\def\OO{{\cal O}}

\def\QQ{{\cal Q}}

\def\TT{{\cal T}}

\def\d{{\partial}}
\newcommand{\pa}{\partial}

\newcommand{\CT}{\mathcal{T}}

\newcommand{\R}{\mathbb{R}}
\newcommand{\T}{\mathbb{T}}

\newcommand{\al}{\alpha}

\newcommand{\nn}{\nonumber}

\newcommand{\spa}{\ , \ \ }

\newcommand{\beq}{\begin{equation}}
\newcommand{\eeq}{\end{equation}}
\newcommand{\ben}{\begin{displaymath}}
\newcommand{\een}{\end{displaymath}}
\newcommand{\beqa}{\begin{eqnarray}}
\newcommand{\eeqa}{\end{eqnarray}}
\newcommand{\bea}{\begin{eqnarray}}
\newcommand{\eea}{\end{eqnarray}}
\newcommand{\bean}{\begin{eqnarray*}}
\newcommand{\eean}{\end{eqnarray*}}
\newcommand{\ba}{\begin{array}}
\newcommand{\ea}{\end{array}}
\newcommand{\bi}{\begin{itemize}}
\newcommand{\ei}{\end{itemize}}
\newcommand{\ie}{{\it i.e.,\,}}
\newcommand{\eg}{{\it e.g.,\,}}

\newcommand{\mc}[1]{\mathcal{#1}}

\newcommand{\lp}{\left(}
\newcommand{\rp}{\right)}

\newcommand{\vep}{\varepsilon}

\newcommand{\si}{{\sigma}}

\newcommand{\mbT}{\mbox{\boldmath$\mathcal T_{\text{tot}}$}}

\numberwithin{equation}{section}

\begin{document}

\begin{titlepage}

\rightline{\vbox{\small\hbox{\tt NORDITA-2011-49} }}
\rightline{\vbox{\small\hbox{\tt CCTP-2011-16}} }

\vskip 3.cm
\begin{center}
\font\titlerm=cmr10 scaled\magstep4
    \font\titlei=cmmi10 scaled\magstep4
    \font\titleis=cmmi7 scaled\magstep4
    \centerline{\titlerm
     Blackfolds in Supergravity and String Theory
      \vspace{0.4cm}}
\vskip 1.cm
{\bf 
Roberto Emparan$^{a,b}$, Troels Harmark$^{c}$,
Vasilis Niarchos$^{d}$,}
{\bf
Niels A. Obers$^{e}$
}
\vskip 0.5cm
\medskip
\textit{$^{a}$Instituci\'o Catalana de Recerca i Estudis
Avan\c{c}ats (ICREA)\\
Passeig Llu\'{\i}s Companys 23, E-08010 Barcelona, Spain}\\
\smallskip
\textit{$^{b}$Departament de F{\'\i}sica Fonamental and
Institut de Ci\`encies del Cosmos, \\
Universitat de Barcelona,
Marti i Franqu{\`e}s 1,
E-08028 Barcelona}\\
\smallskip
\textit{$^{c}$NORDITA,
Roslagstullsbacken 23,
SE-106 91 Stockholm,
Sweden}\\
\smallskip
\textit{$^{d}$Crete Centre for Theoretical Physics,
Department of Physics, University of Crete, 71003}\\
\smallskip
\textit{$^{e}$The Niels Bohr Institute,}
\textit{Blegdamsvej 17, 2100 Copenhagen \O, Denmark}

\vskip .2 in
{\tt emparan@ub.edu, harmark@nordita.org, niarchos@physics.uoc.gr,
obers@nbi.dk}

\end{center}
\vskip 1.5 cm
\centerline{\bf Abstract}
\vskip 0.3cm
\noindent
We develop the effective worldvolume theory for the dynamics of black
branes with charges of the kind that arise in many supergravities and
low-energy limits of string theory. Using this theory, we construct numerous
new rotating black holes with charges and dipoles of D-branes,
fundamental strings and other branes. In some instances, the black holes
can be dynamically stable close enough to extremality. Some of these
black holes, such as those based on the D1-D5-P system, have extremal,
non-supersymmetric limits with regular horizons of finite area and a
wide variety of horizon topologies and geometries. 

\baselineskip 20pt
%

\vskip .5cm \noindent

\vfill
\end{titlepage}\vfill\eject

\setcounter{equation}{0}

\pagestyle{empty}
\small
\tableofcontents
\normalsize
\newpage
\pagestyle{plain}
\setcounter{page}{1}

\section{Introduction}
\label{intro}

In \cite{Emparan:2009cs,Emparan:2009at} we have developed an effective
worldvolume theory, called the blackfold approach, for the dynamics of
black branes on length scales larger than the thickness of the brane.
These initial works focused on the simplest case of vacuum black branes,
but it is natural to extend the approach to black branes with charges.
Such branes play an important role in higher-dimensional supergravity
theories, especially in those that arise in the low energy limit of
string/M-theory. Of particular interest for string theory are black
$p$-branes that carry charges of a Ramond-Ramond field strength
$F_{(p+2)}$. The extremal supersymmetric limit of these black branes is
associated with the supergravity field of (stacks of) D-branes. Many
aspects of the worldvolume dynamics of these (locally) supersymmetric
branes are appropriately captured by the Dirac-Born-Infeld action. The
fundamental strings (F1) and Neveu-Schwarz five-branes (NS5), as well as
the M2 and M5 branes of M-theory often play similar roles. In
supergravity it is natural to consider not only the supersymmetric
branes with these charges, but also their `blackened' versions, with
non-extremal horizons of finite temperature. Thus we seek a theory of
the worldvolume dynamics of black D-branes and other charged black
objects. 

Besides carrying a RR $p$-brane charge, D$p$-branes can also carry `dissolved' in their
worldvolumes other $q$-branes, with $q\leq p$. The
blackfold techniques can naturally accommodate these, by considering
black branes with the charges of $q$-branes, $0\leq q\leq p$, that are
sources of $(q+2)$-form gauge field strengths $F_{(q+2)}$. Then the
worldvolume contains a conserved $q$-brane number
current. Some of these have been studied in \cite{Caldarelli:2010xz}, which develops the
blackfold formalism for branes with only 0-brane and 1-brane charges and
constructs many new classes of charged black holes. It must be borne in
mind that, when considering blackfolds with a spatially compact
worldvolume 
in a globally flat Minkowski
background, the currents that source the field $F_{(q+2)}$ give rise to a
net conserved charge only when $q=0$. When $q\geq 1$ the charge is of
dipole type. So, for instance, out of a black D$p$-brane we 
construct black holes with D$p$ dipole. It is worth noting that the
presence of the latter typically contributes to the stability of the
black hole. In particular, close enough to extremality this charge can
suppress the Gregory-Laflamme-type instability that afflicts neutral
blackfolds.

Quite generally, the blackfold
techniques are the appropriate tool for the study of configurations of
D-branes in the probe approximation, in the case that the D-brane
worldvolume theory has a thermal population of excitations. The
blackfold gives a gravitational description of this thermally excited
worldvolume, with a horizon that on short scales is like that of the
straight black D-brane. Like in the AdS/CFT correspondence, this
gravitational description of the worldvolume theory is appropriate when
there is a stack of a large number of D-branes (although not so large as
to cause a strong backreaction on the background) and the theory is
strongly coupled. Refs.~\cite{Grignani:2010xm} applied these methods to
study a thermal version of the D3-brane BIon.

Since there is a large array of possible $q$-brane charges that a black
$p$-brane can support, for the sake of clarity we confine
ourselves mostly to a selection of simple and potentially relevant
cases. Thus, we begin our analysis in section~\ref{sec:1charge} with the
study of black $p$-folds with only $p$-brane charge. This kind of charge
is not only relevant for (unsmeared) D$p$-branes, but it also presents
some qualitative differences with lower brane charges. The formalism
allows us to obtain, in section~\ref{sec:oddsph}, solutions for thermal
D-branes with a worldvolume whose spatial geometry has the shape of
products of spheres, and in section~\ref{sec:GLstab} to study the
conditions for stability under worldvolume perturbations. The extremal
limit of the solutions exhibits a number of interesting features which
are not captured by the DBI theory.

In section~\ref{sec:qcurrents} we move on to study the more general case
of black $p$-branes with $q$-brane currents, $0\leq q\leq p$. Here our
study of the equations is less detailed and exhaustive, but much of our
analysis is quite general and allows for an arbitrary number of
currents. Again, we find large classes of solutions on products of
spheres. Then we proceed to illustrate the method on specific
two-charge systems: D0-D$p$ and F1-D$p$
(section~\ref{sec:twocharge}) and the extremal D1-D5-P system
(section~\ref{sec:D1D5P}). Our analysis of the latter is concise, to
avoid being repetitious, but its importance comes from the fact that it
gives new extremal (but non-supersymmetric) black holes with a regular
horizon with a wide variety of topologies, and which are potentially
stable. Thus they show how the blackfold techniques can uncover new
classes of black holes that have many of the properties of the solutions
that have become our best laboratory to study black holes in string
theory. Aspects of the string microphysics of our solutions are
addressed in the concluding section~\ref{sec:discuss}.

Our notation follows \cite{Emparan:2009at}, but there are new
features from the inclusion of charge currents that may be worth spelling out: 
\begin{itemize}
\item $\mc Q_q$ is the local charge density of $q$-branes on the
worldvolume of the $p$-brane; $Q_q$ is the total integrated (dipole)
charge on the blackfold. Note that $\mc Q_p=Q_p$.
\item $\Phi_q$ is the local potential conjugate to $\mc Q_q$ on the brane;
$\Phi_H^{(q)}$ is the global potential, conjugate to $Q_q$, for the
black hole. 
\item $h^{(q)}_{ab}$ is the metric on the worldvolume along the $q$-brane
current.
\end{itemize}

\section{Blackfolds with $p$-brane charge}
\label{sec:1charge}

\subsection{Perfect fluids with conserved $p$-brane charge}

We want to describe the dynamics of a perfect fluid that lives in a
$(p+1)$-dimensional worldvolume $\mc W_{p+1}$ and carries a $p$-brane current
\beq
J=Q_p \hat V_{(p+1)}\,.
\eeq
Here $\hat V_{(p+1)}$ is the volume form on $\mc W_{p+1}$. We assume
that this current is conserved (since it sources
a gauge field in the target background spacetime of the brane), 
so the charge must be constant along the worldvolume,
\beq
\d_a Q_p=0\,.
\eeq
Thus $Q_p$ is {\it not} a collective variable of the fluid: there are no
modes in the worldvolume that describe local
fluctuations of this charge\footnote{On a black brane, this
charge can fluctuate on the horizon in directions
transverse to the worldvolume, on the small sphere $s^{n+1}$, but these
are
non-hydrodynamic modes of the black brane that are not
included here.}, so there is no local degree of freedom
associated to it. 

Then the collective coordinates of the fluid are
the same as those for a neutral fluid, namely the local fluid
velocity $u$ and the energy density $\vep$. The presence
of $Q_p$ manifests itself in the equation of state of the fluid, where it
enters as a parameter. 
In addition to the intrinsic variables $\vep$ and $u^a$, the worldvolume geometry is
characterized by the embedding coordinates $X^\mu(\sigma^a)$, which
determine the induced metric $\gamma_{ab}=g_{\mu\nu}\partial_a
X^\mu\partial_b X^\nu$.

The perfect fluid is characterized by its isotropic stress-energy tensor
\beq
\label{perfstress}
T^{ab}=(\vep+P)u^a u^b+P\gamma^{ab}
~,
\eeq
and the equation of state, which will be specified later. Locally it satisfies the
thermodynamic relations
\beq\label{gibbsduhem}
d\vep={\mc T}ds,\qquad
\vep+P={\mc T}s,
\eeq
with $\mc T$ the local temperature and $s$ the entropy density. Note
that $Q_p$, which as we saw is not a local
variable, does not appear here, although we will see later in what sense
it can play a role in the thermodynamics.

The general analysis is very similar to that of perfect neutral fluids:
the fluid equations $D_a T^{ab}=0$ decompose into components parallel (timelike) and
transverse (spacelike) to $u^a$. The former gives the
energy continuity equation, which for a perfect fluid, and using
\eqref{gibbsduhem}, is equivalent to the conservation of entropy
\beq\label{entcont}
D_a(su^a)=0\,.
\eeq
The (spacelike) Euler force equations relate the fluid acceleration along the
worldvolume, $\dot u^a=u^b D_b u^a$, to the pressure gradient and, through
the thermodynamic identity $d P=sd\mc T$, to the temperature gradient so that 
\beq\label{euler}
(\gamma^{ab}+u^a u^b)(\dot u_b+\partial_b\ln\mc T)=0\,.
\eeq

The worldvolume of the fluid is also dynamical since we allow
fluctuations of its embedding $X^\mu(\sigma^a)$ in the background. Its
shape is captured
by the first
fundamental form $h^{\mu\nu}=\gamma^{ab}\partial_a X^\mu\partial_b
X^\nu$ and the extrinsic curvature tensor
$K_{\mu\nu}{}^\rho=h_\mu{}^\lambda h_\nu{}^\sigma\nabla_\sigma
h_\lambda{}^\rho$, with mean curvature vector
$K^\rho=h^{\mu\nu}K_{\mu\nu}{}^\rho$.
This elastic dynamics is captured by
Carter's equations \cite{Carter:2000wv}
\beq\label{carter1}
K_{\mu\nu}{}^{\rho}T^{\mu\nu}=\frac1{(p+1)!}\perp^\rho{}_\sigma
J_{\mu_0\dots\mu_p}F^{\mu_0\dots\mu_p\sigma}\,.
\eeq
Here $\perp^\mu{}_\nu=\delta^\mu{}_\nu-h^\mu{}_\nu$ is the projector
orthogonal to the worldvolume, and
for generality we include a force from the coupling of the current $J$ to a
background $(p+2)$-form field strength $F_{[p+2]}$. However, in this
article we set this background field to zero. Then, for
the perfect fluid stress tensor \eqref{perfstress} these equations become
\beq\label{carter2}
-P K^\rho=\perp^\rho{}_\mu\, s{\mc T}\dot u^\mu\,.
\eeq
This equation determines the acceleration of the fluid along directions orthogonal to
the worldvolume as an effect of forcing by the extrinsic curvature. 

\subsubsection{Stationary solutions}

In order to proceed further, we restrict ourselves to stationary
configurations, for which the fluid velocity is aligned with a Killing
vector $k$ along the worldvolume. Moreover, we assume that
$k=k^\mu\partial_\mu$ generates
isometries not only of the worldvolume but also of the background spacetime. Thus
\beq
\label{stat1aa}
u=\frac{k}{|k|}~, \qquad \nabla_{(\mu}k_{\nu)}=0\,,
\eeq
and $\dot u_\mu=\partial_\mu\ln|k|$.
This determines completely the solution to the intrinsic fluid equations,
since the local temperature is obtained by simply redshifting the global
temperature $T$, which is uniform on the worldvolume,
\beq\label{Tshift}
\mc T(\sigma^a)=\frac{T}{|k|}\,.
\eeq
Since now $s\mc T \dot u_\mu=-s\partial_\mu\mc T=-\partial_\mu P$, the
extrinsic equation \eqref{carter2} is
\beq\label{exteqs}
K^\rho=\perp^{\rho\mu}\partial_\mu\ln(- P)
\eeq
($P$ is negative in the examples we consider). This equation can be
obtained by extremizing the action
\beq\label{actQ}
\tilde I=\int_{\mc W_{p+1}}d^{p+1}\sigma \sqrt{-\gamma}\;P
\eeq
for variations of the brane embedding among stationary fluid
configurations with the same $Q_p$.

In order to compute the physical magnitudes of
the brane configuration for any given embedding, we must first specify
how $k$ is related to the background Killing vector $\xi$ that defines
unit-time translations at asymptotic infinity. Without loss of
generality we write
\beq\label{kxichi}
k=\xi+\Omega \chi
\eeq
where $\chi$ is a spacelike Killing vector, typically a generator of
rotations, orthogonal to $\xi$, and $\Omega$ a constant that gives the
(angular) velocity relative to orbits of $\xi$. We assume that on the worldvolume $\mc
W_{p+1}$ the vector $\xi$ is orthogonal to spacelike hypersurfaces $\mc
B_{p}$, with unit normal\footnote{It is easy, but slightly cumbersome,
to extend this to the case where the normal to $\mc
B_{p}$ is not parallel to the generator of asymptotic time translations
\cite{Camps:2008hb}.}
\beq\label{nxi}
n^a=\frac1{R_0}\xi^a|_{\mc W_{p+1}}\,.
\eeq
The factor $R_0(\sigma^a)=-n^b \xi_b$ measures the local gravitational
redshifts between points on $\mc W_{p+1}$ and asymptotic infinity
(where $\xi^2\to -1$)\footnote{For instance this factor is non-trivial
for blackfolds in AdS backgrounds \cite{Caldarelli:2008pz}.}.
It is also convenient to introduce the rapidity $\eta$ of the
fluid velocity relative to the worldvolume time generated by $n^a$,
such that
\beq\label{naua}
-n^a u_a =\cosh\eta\,.
\eeq
If we denote $R^2=\left.|\chi^a\chi_a\right|_{\mc W_{p+1}}$
then
\beq
\tanh\eta=\frac{\Omega R}{R_0}\,.
\eeq

Integrations on the worldvolume $\mc W_{p+1}$
reduce, over an
interval $\Delta t$ of the Killing time generated by $\xi$, to integrals over $\mc
B_{p}$ with measure $dV_{(p)}$ as
\beq\label{intWB}
\int_{\mc W_{p+1}} d^{p+1}\sigma \sqrt{-\gamma}\,(\dots)=\Delta t\int_{\mc
B_{p}}R_0dV_{(p)}\,(\dots)\,.
\eeq

The mass, angular momentum and entropy are now obtained in a
conventional manner as integrals over $\mc
B_{p}$,
\beqa\label{MJS}
M=\int_{\mc B_{p}}dV_{(p)} T_{ab}n^a\xi^b\,,\qquad 
J=-\int_{\mc B_{p}}dV_{(p)} T_{ab}n^a\chi^b\,,\qquad
S=-\int_{\mc B_{p}}dV_{(p)} s u_a n^a\,.
\eeqa
For the stress tensor \eqref{perfstress}, and using \eqref{nxi} and
\eqref{naua}, it is easy to find the general expressions for the
integrands of these
magnitudes in terms of $\vep$, $P$, $s$ and $\eta$.

\subsubsection{Action principles: fixed charge and fixed potential}

Contracting \eqref{perfstress} with $n_a k_b$ and using \eqref{gibbsduhem} we find
\beq\label{stressnk}
T_{ab}n^a(\xi^b+\Omega\chi^b)+T s\, n^a u_a=n^a k_a P=-R_0 P\,.
\eeq
This can be regarded as a version of \eqref{gibbsduhem} where we take
into account fluid motion and gravitational redshifts.
Integrating over $\mc B_p$ and using \eqref{intWB} we obtain the action
\eqref{actQ} in the form
\beq
\tilde I=-\Delta t(M-\Omega J-TS)\,.
\eeq
We can rotate to Euclidean time, $it\to \tau$, with periodicity
$\Delta\tau=\beta=1/T$, and find that the Euclidean action, $i\tilde I\to -\tilde I_E$ is
\beq
\tilde I_E=\beta(M-\Omega J-TS)\,,
\eeq
\ie the thermodynamic potential at constant $T$ and $\Omega$, with $Q_p$ fixed.

Since $T$ and $\Omega$ are integration constants, the extrinsic
equations \eqref{exteqs} are equivalent to requiring that
\beq
dM=TdS+\Omega d J\qquad \mathrm{(fixed~}Q_p)\,.
\eeq
This is, we extremize the action for variations among
configurations with the same value of $T$, $\Omega$, and the charge $Q_p$.

Note that $Q_p$ is not a worldvolume density but a global quantity for the
fluid on the brane. Even if there is no local, intrinsic fluid degree of
freedom on the worldvolume associated to this charge, it is one of the
conserved total charges of the configuration, together with $M$ and $J$.
Thus we should be able to consider variations among stationary
fluid brane configurations with different charges. This requires
the existence, for stationary branes, of a global
potential $\Phi_H^{(p)}$ conjugate to $Q_p$. In order to identify it, note that
the local equation of state that relates $\vep$ to $s$ does also depend on
$Q_p$ as a parameter that specifies the entire fluid, $\vep(s;Q_p)$. Thus on
the fluid worldvolume we can introduce a local potential $\Phi_p(\sigma^a)$
as
\beq
\Phi_p(\sigma^a)=\frac{\partial\vep(s;Q_p)}{\partial Q_p}\,.
\eeq 
Since $Q_p$ couples to the field strength $F_{[p+2]}$ of the background,
this $\Phi_p$ actually
corresponds to the potential of $F_{[p+2]}$ on the worldvolume (for a black
brane,
this quantity is well defined without any arbitrary additive
constants \cite{Emparan:2004wy,Copsey:2005se}).
We introduce 
the global potential
\beq\label{Phidef}
\Phi_H^{(p)}=\int_{\mc B_p} dV_{(p)} R_0 \Phi_p(\sigma^a)
\eeq
by integrating over spatial directions of the worldvolume taking
into account the local redshift $R_0$ relative to infinity. 

We can now reformulate the variational principle for stationary
solutions using this potential.
Locally, we introduce the
Gibbs free energy density
\beq\label{gibbs}
\mc G=\vep -\mc T s -\Phi_p Q_p=-P-\Phi_p Q_p
\eeq
which, for hydrodynamic fluctuations for which $Q_p$ is necessarily
constant along the worldvolume, satisfies\footnote{Note that $\mc T$ is
a true independent local variable of the fluid, which characterizes
local fluctuations in the energy density of the fluid, but
$\Phi_p(\sigma^a)$ is not: it measures the response of the fluid to an
external source that {\it globally} changes the fluid's $Q_p$, which
remains constant over the worldvolume. So, for the
intrinsic dynamics of the fluid, this `local grand-canonical ensemble'
is a phony one: $\mc G$ is only a function of $\mc T$ since $\Phi_p$ is
determined by the condition that $Q_p$ remains constant.}
\beq
d\mc G=-sd\mc T -Q_pd\Phi_p=-dP-Q_pd\Phi_p\,.
\eeq

The extrinsic equations
\eqref{exteqs} can now be written as
\beq\label{exteqs2}
(\mc G +\Phi_p Q_p)K^\rho=\perp^{\rho\mu}(\partial_\mu\mc
G+Q_p\partial_\mu\Phi_p).
\eeq

Let us consider variations of the worldvolume embedding where, instead
of $Q_p$, we keep
constant the potential $\Phi_H^{(p)}$ in \eqref{Phidef}. To achieve this,
the local potential $\Phi_p(\sigma^a)$ must vary in such a
way that
\beq
\perp^{\rho\mu}\partial_\mu\Phi_p(\sigma^a)=\Phi_p(\sigma^a)K^\rho
\eeq
(see eq.~(A.40) of \cite{Emparan:2009at}). Then, the extrinsic equations \eqref{exteqs2} take the form
\beq
K^\rho=\perp^{\rho\mu}\partial_\mu\ln \mc G
\eeq
which we can derive by extremizing the action
\beq\label{actPhi}
I=-\int_{\mc W_{p+1}}d^{p+1}\sigma \sqrt{-\gamma}\;\mc G
\eeq
for variations of the embedding among stationary fluid configurations
where now $T$, $\Omega$ and $\Phi_H^{(p)}$ are kept fixed. By
integrating \eqref{gibbs} and going to Euclidean time 
we recover the
expected Legendre-transform-type relations
\beqa
I_E[T,\Omega,\Phi_H^{(p)}]&=&\beta(M-\Omega J-TS-\Phi_H^{(p)} Q_p)\nonumber\\
&=&\tilde I_E[T,\Omega,Q_p]-\beta\Phi_H^{(p)} Q_p\,,
\eeqa
and $\beta\Phi_H^{(p)}=\partial \tilde I_E/\partial Q_p$ which, consistently, justifies
the way we have defined $\Phi_H^{(p)}$ in \eqref{Phidef} in terms of
worldvolume quantities. The extrinsic equations are now equivalent to the complete
form of the global first law,
\beq
dM=TdS+\Omega dJ+\Phi_H^{(p)} dQ_p\,.
\eeq

\subsection{Blackfold effective fluid}
\label{geneqs}

In order to apply the previous formalism to a specific kind of brane we
need to know the equation of state of the effective fluid that
lives in its worldvolume. Here we consider the charged
dilatonic black $p$-branes solutions of the
action
\beq
\label{IIMaa}
I=\frac{1}{16\pi G} \int d^D x~\sqrt{-g}
\left( R-\frac{1}{2}(\d \phi)^2-\frac{1}{2(p+2)!} e^{a\phi}F^2_{(p+2)} \right)
\eeq
with general parameters $D$, $p$. Defining
\beq
\label{IIMab}
n=D-p-3\,, \qquad a^2=\frac{4}{N}-\frac{2(p+1)n}{D-2}
\eeq
the flat black $p$-brane solution reads
\beq
\label{IIMaca}
ds^2=H^{-\frac{Nn}{D-2}}\left( -f dt^2+\sum_{i=1}^p dx_i^2 \right)
+H^{\frac{N(p+1)}{D-2}}\left(f^{-1} dr^2+r^2 d\Omega^2_{n+1} \right)
\,,
\eeq
\beq
\label{IIMac}
e^{2\phi}=H^{aN}~, ~~ A_{(p+1)}=\sqrt{N}\coth \alpha (H^{-1}-1)
dt\wedge dx_1\wedge\cdots \wedge dx_p
\,,
\eeq
with
\beq
\label{IIMad}
H=1+\frac{r_0^n \sinh^2\alpha}{r^n}
\,, \qquad f=1-\frac{r_0^n}{r^n}
\,.
\eeq
 Note that
since it must
be that $a^2\geq 0$ (otherwise the dilaton would be a
ghost), the parameter $N$ cannot be arbitrarily large but is bounded by
\beq\label{Nbound}
N\leq 2\left(\frac1n +\frac1{p+1}\right)\,.
\eeq
In string/M-theory, $N$ is typically an integer up to 3 (when $p\geq 1$) that
corresponds to the number of different types of branes in an intersection. 
The D/NS-branes in type II string theory are obtained for $D=10$, $N=1$, $p=0,\dots,6$.
The M-branes in M-theory arise for $D=11$, $N=1$, $p=2,5$.

It is straightforward to compute the properties of the effective stress tensor of the
fluid as functions of $r_0$ and $\alpha$,
\beqa\label{bffluid1}
\varepsilon=\frac{\Omega_{(n+1)}}{16\pi G}r_0^n\lp n+1+n N\sinh^2\al\rp,\qquad
P=-\frac{\Omega_{(n+1)}}{16\pi G}r_0^n\lp1+n N\sinh^2\al\rp,\label{bffluid2}
\eeqa
\beq
{\mathcal T}=\frac n{4\pi r_0\lp\cosh\al\rp^{N}} ,\qquad
s=\frac{\Omega_{(n+1)}}{4G}r_0^{n+1}\lp\cosh\al\rp^{N},
\label{Ts}\eeq
and 
\beq
Q_p=\frac{\Omega_{(n+1)}}{16\pi G}n\sqrt{N}r_0^n\sinh\al\cosh\al\,,\qquad
\Phi_p
=\sqrt N\tanh\al\,.\label{Qbrane}\eeq
The potential $\Phi_p$ is measured from the difference between the values of $A_{(p+1)}$
at the horizon and at $r\to\infty$ in \eqref{IIMac}. The Gibbs free
energy density takes a particularly simple form
\beq
\label{Gbrane}
\mc G=\frac{\Omega_{(n+1)}}{16\pi G}r_0^{n}\,.
\eeq

If we eliminate $r_0$ and $\alpha$ from these expressions we obtain the
equation of state that relates $\vep$ to $s$ and $Q_p$. However it is
convenient to retain $r_0$ and $\alpha$, not only as a useful
parametrization, but also
 for relating 
to the geometrical properties of the black hole solutions: $r_0$
and $\alpha$ determine the geometry in the region close to the black
brane. Thus they give information about
the short-scale structure of the black holes. 

A number of simple relations can be found among physical quantities which,
although not independent of the local thermodynamic relations
and the equation of state, are illustrative. The relation
\beq\label{bfeos}
\vep=-(n+1)P-n\Phi_p Q_p\,,
\eeq
says that the effect of the charge is to provide an additional
tension on the worldvolume. This also becomes apparent in the extrinsic equation
\beq\label{exteqn1}
K^\rho=\frac{n \mc T s}{\mc T s+n\Phi_p Q_p}\perp^\rho{}_\mu \dot u^\mu
=\frac{n}{1+nN\sinh^2\alpha}\perp^\rho{}_\mu \dot u^\mu\,,
\eeq
according to which the acceleration required in order to balance a given
extrinsic curvature increases when charge is present.

An equation equivalent to \eqref{bfeos} gives the Gibbs free energy density as
\beq\label{GibbsTs}
\mc G=\frac1n \mc T s\,,
\eeq
which, taking $\beta=1/T$, implies that
\beq\label{acts}
I_E=\frac1{n}S\,.
\eeq
Using relations analogous to \eqref{stressnk}, we can easily prove that
\beq\label{presmarr}
(D-3)M-(D-2)(TS+\Omega J)-n \Phi_H^{(p)} Q_p=\mbT
\eeq
where the total tensional energy is
\beq\label{ttot}
\mbT=-\int_{\mc B_{p}}dV_{(p)}R_0(\gamma^{ab}+n^a n^b)T_{ab}\,.
\eeq
In the case of a Minkowski background the black hole will be
asymptotically flat and it must satisfy a Smarr relation. This is
precisely of the form of eq.~\eqref{presmarr} with $\mbT=0$. This
condition is therefore necessary for equilibrium in these backgrounds,
and must follow after solving the extrinsic equations. We will see how
it can be used for an efficient determination of the equilibrium value
of the rapidity $\eta$.

Observe that the effective stress-energy tensor can be
written as
\beqa\label{gex}
T_{ab}=\mc T s\lp u_a u_b-\frac1{n}\gamma_{ab}\rp
-\Phi_p Q_p\,\gamma_{ab}
\eeqa
We will see later that this form is generic for charged blackfold
fluids. It suggests that the stress tensor has a brane-tension component
$\propto -\Phi_p Q_p$, and a thermal component $\propto \mc T s$. However,
one must bear in mind that this is not a decomposition into ground-state
and thermal-excitation energies: the potential $\Phi_p$ changes depending
on the temperature of the system. The actual ground state,
discussed below, has the uniform value $\Phi_p=\sqrt{N}$ at zero temperature.

Finally, let us recall that the charge $Q_p$ is a dipole charge of the
black hole constructed as a blackfold with compact $\mc B_p$. This
is not a conserved asymptotic charge, and the definition of its
conjugate potential $\Phi_H^{(p)}$ is somewhat subtle. Global issues such as
those found in refs.~\cite{Emparan:2004wy,Copsey:2005se} are presumably relevant in our
constructions too (\eg at the center of the $S^p$ in the configurations
of sec.~\ref{sec:oddsph} below, when $n=1$), but we do not address them
here. The fact that our definition of $\Phi_H^{(p)}$ leads to the expected
form of black hole thermodynamic relations gives us confidence that
these issues can be satisfactorily handled.

\subsection{No boundaries}

In contrast to neutral blackfolds, and also to blackfolds with lower-form
currents, blackfolds with $p$-brane charges do not admit open
boundaries: the charge $Q_p$ would not be conserved at them. While it is
possible to consider that the blackfold ends on another brane that
carries the requisite charge (and if this is a black brane, then the
boundary will be one where $R_0=0$), we will not consider such situations in
this article.

This rules out, in particular, the existence of blackfold disk and ball
solutions of the type obtained in
\cite{Emparan:2009cs,Emparan:2009vd,Caldarelli:2010xz}.

\subsection{Extremal limits \label{sec:extremal}}

The black branes in \eqref{IIMaca} have extremal limits in which the
horizon becomes degenerate. We can therefore consider
blackfolds in which the horizon approaches this limit uniformly over all
of $\mc B_{p}$. This is in contrast with the situation in
\cite{Emparan:2009vd,Caldarelli:2010xz} where
the extremal limit was reached locally at the boundaries of $\mc
B_{p}$, while the rest of $\mc B_{p}$ remained non-extremal. In the
present case there can be no boundaries so this possibility is absent.

For most of the black branes in \eqref{IIMaca} the horizon becomes
singular in these extremal limits, but these are `good singularities' in
the sense explained in \cite{Gubser:2000nd}: we can approach them as
limits of solutions with smooth non-degenerate horizons, so the
singularity can be regarded as the result of integrating out
short-distance degrees of freedom that can be thermally excited. Thus
these solutions have sensible physical interpretations. The blackfold
method can be applied to them, even if the intrinsic dynamics is not a
fluid dynamics anymore.

\subsubsection{Dirac branes}\label{subsubsec:dirac}

The straightforward limit to an extremal solution is obtained by
taking
\beq\label{extlim1}
r_0\to 0\,,\qquad \alpha\to\infty
\eeq
while keeping the charge $Q_p$ fixed. In the limit, $\mc T s\to 0$ and $\Phi_p\to\sqrt{N}$ so
\beq
\vep=-P=\sqrt{N} Q_p\,.
\eeq
Since $Q_p$ is constant on the worldvolume, it
follows that the energy density and pressure must be uniform too. The
stress tensor
\beq\label{extTab}
T_{ab}=P\gamma_{ab}
\eeq
is that of a brane with uniform tension, whose
action is proportional to the volume of $\mc W_{p+1}$, as studied by Dirac.
In the cases where
the charge is that of a D-brane, this is the same as the
Dirac-Born-Infeld action with
vanishing Born-Infeld gauge fields. Thus we refer to them as Dirac branes.

Since these branes have local Lorentz invariance on the worldvolume, the
velocity field is pure gauge and is not a physical collective mode
anymore.
The only dynamics is extrinsic, and in the absence of external form
fields, is given by the minimal-surface equations 
\beq 
K^\rho=0\,. 
\eeq
The absence of compact embedded minimal surfaces in Euclidean space
implies that in a flat Minkowski background there are no solutions to
these equations that would correspond to static extremal black holes
with $p$-brane dipole. This is in the same spirit as the no-dipole-hair
theorem of \cite{Emparan:2010ni}, which forbids a large class of non-extremal static
dipole black holes.

\subsubsection{Null-wave branes}
\label{subsubsec:momwaves}

A more interesting limit results if, in addition to \eqref{extlim1}, we scale
the velocity field components to infinity in such a way that
$r_0^{n/2}u^a$ remains finite. Let us introduce then a `momentum
density' ${\mc K}$ and a vector $l$ that remain finite in the limit,
\beq
\lp \frac{\Omega_{(n+1)}}{16\pi G}n r_0^n\rp^{1/2}u^a ={\mc K}^{1/2} l^a\,.
\eeq
Thus, when $r_0\to 0$,
\beq
l_a l^a=0
\eeq
so $l$ is a lightlike vector: the fluid on the brane is boosted to the
speed of light. The stress tensor becomes
\beq\label{nulltab}
T_{ab}={\mc K}\, l_a l_b -\sqrt{N}Q_p\,\gamma_{ab}\,.
\eeq
Thus, in addition to the tensional, ground state component $\propto Q_p$,
the brane supports a
null momentum wave with momentum density ${\mc K}$, without breaking
locally the translational invariance along the wave. This stress tensor
cannot be obtained from the DBI action. Since $\vep+P=\mc T s$
vanishes in the limit, we
are also outside the
realm of proper fluid dynamics.

The dynamics of these extremal null-wave branes is in some respects simpler than for
non-extremal branes, so we will briefly describe some aspects of it. As is customary
when dealing with null congruences generated by a vector $l^a$, we
introduce another null vector $\bar l^a$ such that
\beq
\bar l_a \bar l^a=0\,,\qquad \bar l_a l^a=-1\,,\qquad 
l^aD_a\bar l^b=0\,,
\eeq
and the projector onto the $(p-1)$-dimensional orthogonal space
\beq
q_{ab}=\gamma_{ab}+l_a\bar l_b+l_b\bar l_a\,.
\eeq
The intrinsic equations $D_a T^{ab}=0$ are then projected along
$l^a$, $\bar l^a$, and $q_{ab}$. The equation $l_bD_a T^{ab}=0$ is automatically
satisfied, while $\bar l_bD_a T^{ab}=0$ implies
\beq
D_a ({\mc K} l^a)=0
\eeq
so ${\mc K} l^a$ is a conserved momentum current. Finally, $q_{cb}D_a T^{ab}=0$
gives
\beq
l^a D_a l^b=0\,,
\eeq
so $l^a$ generates geodesics along the worldvolume. 
However, these need
not be geodesics of the background, and indeed the acceleration of $l$ transverse
to the worldvolume appears in the extrinsic equations
\beq
{\mc K}{\perp^\rho}_\mu l^\nu\nabla_\nu l^\mu=\sqrt{N}Q_p K^\rho
\eeq
as balancing the tension from extrinsic curvature. 

Since we are not dealing with a worldvolume fluid, the conditions for
absence of dissipative effects in stationary solutions are less clear
than in non-extremal blackfolds. In particular, it does not seem to be
required that the the null vector $l^a$ is a Killing vector on $\mc
W_{p+1}$ --- although this is presumably a necessary requirement for horizon
regularity when the
extremal brane has a horizon with finite area \cite{Horowitz:2004je}. 

Nevertheless, there are some general conditions that are satisfied in Minkowski
backgrounds, with $R_0=1$. As we have seen, stationary equilibrium in
these cases requires that the tensional energy \eqref{ttot} vanish.
Applied to \eqref{nulltab}, and taking $l^a$ to be normalized
so that $n^a l_a=-1$, the condition $\mbT=0$ relates the
averaged momentum density to the
charge as
\beq\label{KQ}
\frac1{V_{(p)}}\int_{\mc B_p}dV_{(p)}\mc K =p\sqrt{N} Q_p \,.
\eeq
Furthermore, since the mass \eqref{MJS} is
\beq\label{MiKQ}
M=\int_{\mc B_p}dV_{(p)}\mc K +\sqrt{N} V_{(p)} Q_p\,,
\eeq
we can derive that
\beq\label{MQK}
\frac1{p+1}M= \sqrt{N}V_{(p)} Q_p=\frac1p \int_{\mc B_p}dV_{(p)}\mc K\,.
\eeq
These expressions are of the kind previously obtained in \cite{Caldarelli:2010xz}, and
they show how the total energy $M$ of the brane in equilibrium is
virialized between kinetic energy and potential (charge-tensional)
energy. For fundamental strings ($p=1$, $N=1$), \eqref{KQ} says that at
equilibrium the momentum and winding numbers must be equal \cite{BlancoPillado:2007iz}.

The angular momentum in the direction of the
rotational vector $\chi^a$ is
\beq
J=\int_{\mc B_p}dV_{(p)}\mc R \mc K
\eeq
where
\beq
\mc R=\chi^a l_a\,
\eeq
has the interpretation of `lever-arm' radius for the momentum.
In cases where this radius is constant over all of $\mc B_p$, like in the round odd-spheres
discussed later, we obtain
\beq\label{MQJ}
\frac1{p+1}M= \sqrt{N}V_{(p)} Q_p=\frac1p \frac{J}{\mc R}\,.
\eeq
Provided that $\Omega=1/\mc R$, this also follows from the
extremal limit $TS\to 0$ of the Smarr relation \eqref{presmarr} and
\eqref{acts}. 

Eqs.~\eqref{MQK} and \eqref{MQJ} are correct only to leading order in
the blackfold expansion and in general receive corrections at the next
orders, since the gravitational and gauge self-interaction of the
brane gives rise to Newtonian and Coulombian potential energies that
renormalize the mass. This effect has been studied in
\cite{BlancoPillado:2007iz,Emparan:2008qn} using the exact black ring solutions of
\cite{Emparan:2004wy,Elvang:2004xi}.

\subsection{Stress-energy of excitations. Near-extremal $p$-branes.}

As we have seen, the ground state corresponds to
the extremal limit with stress-energy tensor \eqref{extTab}. So the
stress tensor for excitations above the ground state of a $p$-brane with
given charge $Q_p$ is
\beqa
T^\mathrm{(exc)}_{ab}&=&T_{ab}-T^\mathrm{(ground)}_{ab}
=T_{ab}+\sqrt{N}Q_p\gamma_{ab}\nn\\
&=&\mc T s\lp u_a u_b+\lp \frac{N\Phi_p}{\sqrt{N}+\Phi_p}-\frac1{n}\rp\gamma_{ab}\rp\,.
\eeqa
When the system is near extremality, $\Phi_p\simeq \sqrt{N}$ and
\beq\label{Tabexc}
T^\mathrm{(exc)}_{ab}\simeq \mc T s\lp u_a u_b+\lp \frac{N}{2}-\frac1{n}\rp\gamma_{ab}\rp\,.
\eeq
This can be regarded as the stress-energy of the thermal gas of
excitations of the worldvolume theory on a stack of branes, in the regime
(typically at strong coupling) where this theory is appropriately
described in terms of a gravitational dual. This stress tensor is
traceless when $N/2 -1/n=1/(p+1)$, which implies that the dilaton
coupling $a$ vanishes.
Well-known instances of this are the
D3, M2, M5, all with $N=1$; strings in six dimensions with
$N=2$ (\eg D1-D5); and strings in five dimensions, with
$N=3$ (\eg M5$\perp$M5$\perp$M5).

If we extract an energy density and pressure of excitations from this
stress-energy tensor, then, near extremality we find that whenever $Nn>2$ we have
$dP^\mathrm{(exc)}/d\vep^\mathrm{(exc)}>0$ so the speed of sound is real
and the fluid is stable. This includes all the non-dilatonic black branes
mentioned above. We will return to this issue of stability in more generality below.

\section{Odd-sphere blackfolds}
\label{sec:oddsph}

The simplest solutions to the equations of the previous section that
have spatially compact worldvolume are odd-spheres,
\beq
\mc B_{p}=S^p\,,\qquad p=2m+1
\eeq
so that the horizon topology is $ S^{p} \times s^{n+1}$. In the
Minkowski background the $S^p$ are embedded in a $\R^{p+1}$ subspace
\beq
ds^2=dr^2+r^2d\Omega_{(p)}^2
\eeq
as surfaces at $r=R$.
In principle one may also allow spheres $ S^{p}$ which are not round, but the
solutions can be obtained algebraically only when the radius is
constant and the angular momentum is aligned along a diagonal of the
Cartan subgroup of rotations of $S^{p}$. That is, the angular
velocities along the Cartan generators $\partial/\partial\phi_i$ are all equal
and
the Killing vector
$\chi$ is
\beq\label{chiphi}
\chi=\frac{\partial}{\partial\phi}=\sum_{i=1}^{m+1}\frac{\pa}{\pa\phi_i}\,.
\eeq
Then
\beq
u=\frac1{\sqrt{1-\Omega^2 R^2}}\lp\frac{\partial}{\partial
t}+\Omega\frac{\partial}{\partial \phi}\rp\,.
\eeq
Since in this case $|k|$ is homogeneous on $S^p$, so is
the entire solution, and in particular
 $r_0$ and $\alpha$. 

The simplest way to solve the
extrinsic equations is
by requiring that $\mbT=0$: since all fields
are constant on $S^p$ we need only solve the algebraic equation
\beq
\lp\gamma^{ab}+\delta_t^a\delta_t^b\rp T_{ab}=0\,.
\eeq
This is particularly simple in terms of
the rapidity $\eta$ introduced in \eqref{naua}, so that
$\tanh\eta=\Omega R$. One obtains
\beq\label{soleta}
\sinh^2\eta=\frac{p}{n}(1+nN\sinh^2\alpha)\,.
\eeq
In this form it is apparent that the result from the exact dipole ring
solution of \cite{Emparan:2004wy}, with $p=1=n$, is correctly reproduced. One easily
sees that \eqref{soleta} actually solves \eqref{exteqn1}, since
$K^r=-p/R$ and $\dot u^r=-R^{-1}\sinh^2\eta$. It is only a little more
laborious to verify that this solution extremizes either of the actions
\eqref{actQ}, \eqref{actPhi} with fixed $Q_p$ and $\Phi_H^{(p)}$, respectively. 
The solution determines the radius of the sphere to be
\beq
\label{Roddsphere1}
R=\frac1{\Omega}\sqrt{\frac{p(1+nN\sinh^2\alpha)}{n+p(1+nN\sinh^2\alpha)}}\,.
\eeq
However, note that when expressed in terms of $T$ and $\Phi_H^{(p)}$, the
right hand side of this equation depends on $R$.

Given the homogeneity of the blackfold solution, the integrals
\eqref{MJS}, \eqref{Phidef} that give $M$, $J$, $S$ and $\Phi_H^{(p)}$,
are obtained by simply multiplying the corresponding energy densities by
the volume $V_{(p)} = R^p \Omega_{(p)}$ of a round $p$-sphere with
radius $R$. We present these results in appendix~\ref{app:STMBF}.

When the angular velocities along the Cartan generators are not all
equal, we expect the spheres to be distorted and have longer
circumference along the directions of larger rotation velocity. Compared
to the neutral case, the additional tension from the charge
increases the rigidity of the brane and
will oppose this distortion. Presumably there exists a bound on
the size of the ratio between different angular velocities.

\paragraph{Extremal limit.}

Eq.~\eqref{soleta} clearly shows that the extremal limit
$\alpha\to\infty$ for non-trivial brane solutions (with finite non-zero radius)
is also a limit in which the local boost becomes lightlike,
$\eta\to\infty$. Thus we obtain branes with a null momentum wave on
them, and $\Omega R=1$. In this limit
we can use \eqref{MQJ} and
write $R=\mc R$ as
\beq\label{RJM}
R=\frac{p+1}{p}\frac{J}{M}
\eeq
and since the spatial volume is $V_{(p)}=\Omega_{(p)}R^p$, find a relation
between $M$, $Q_p$, $J$,
\beq\label{QMJ}
Q_p=\frac{1}{\sqrt{N}\,(p+1)\Omega_{(p)}}\lp\frac{p}{p+1}\rp^{p}\frac{M^{p+1}}{J^p}\,.
\eeq
For fixed $M$, this is a curve in the $(Q_p,J)$ plane that determines
the upper bounds on the values that the non-extremal solutions can
have: it gives the maximum value of $J$ for given $M$ and $Q_p$,
and the maximum charge $Q_p$ for given $M$ and $J$ (and the minimum
allowed $M$ for given $J$ and $Q_p$). It is interesting that it depends on
$p$ but not on $n$. This relation is correctly
reproduced for the exact extremal dipole ring solutions ($p=1$) of
\cite{Emparan:2004wy} in the limit of large
ring radius.

\subsection{Products of odd-spheres \label{sec:Dpoddsphere}}

The previous solutions can easily be generalized to products of round odd-spheres,
$\mc B_p=\prod_{I=1}^\ell S^{p_I}$, with odd $p_I$ and $\sum_{I=1}^\ell p_I=p$,
embedded as the surfaces $r_I=R_I$ in
\beq\label{prodS}
ds^2=\sum_{I=1}^\ell \left( dr_I^2+r_I^2d\Omega_{(p_I)}^2\right)\,.
\eeq
A particularly simple case are the tori $\T^p$, where $p_I=1$. Note that
the number of spheres is limited by
\beq\label{ellbound}
\ell\leq n+2\,.
\eeq

The velocity is taken along $\chi=\sum_{I=1}^\ell \Omega_I \partial_{\phi_I}$, where
$\partial_{\phi_I}$ are generators of diagonals of the Cartan subgroups of
$SO(p_I+1)$. Then
\beq\label{uoddsph}
u=\cosh\eta\lp\frac{\partial}{\partial
t}+\sum_{I=1}^\ell\Omega_I\frac{\partial}{\partial \phi_I}\rp
\eeq
where $\eta$ is the total rapidity so that
\beq
\tanh^2\eta=\sum_{I=1}^\ell\Omega_I^2 R_I^2\,.
\eeq 

The extrinsic equilibrium, \eqref{exteqn1}, now involves more than one
equation so it cannot be obtained by simply setting $\mbT=0$. Instead we obtain it
from
\beq\label{Kri}
K^{r_I}=-\frac{p_I}{R_I}\,,\qquad 
\dot u^{r_I}=-\Omega_I^2 R_I\cosh^2\eta\,.
\eeq

Eqs.~\eqref{exteqn1} are now solved by
\beq
\Omega_I R_I=\sqrt{\frac{p_I}{p}}\tanh\eta
\eeq
where $\sinh^2\eta$ is given by
the same expression as in \eqref{soleta}\,.
More explicitly, 
\beq\label{RIOmIal}
R_I=\frac{1}{\Omega_I}\sqrt{\frac{p_I(1+ n N\sinh^2\alpha)}{n+p(1+ n
N\sinh^2\alpha)}}\,.
\eeq
Note that there is a
larger component of the velocity on spheres $S^{p_I}$ of higher
dimensionality --- intuitively, the velocity must counterbalance the
tension in more directions.

In the extremal limit in which the velocity becomes null, we have
\beq\label{OmRIext}
\Omega_I R_I =\sqrt{\frac{p_I}{p}}
\eeq
and the null vector  is
\beq\label{lodd}
l=\frac{\partial}{\partial
t}+\sum_{I=1}^\ell\sqrt{\frac{p_I}{p}}\frac{1}{R_I}\frac{\partial}{\partial
\phi_I}\,.
\eeq

Given the homogeneity on $\mc B_p$, eq.~\eqref{KQ} becomes
\beq
\mc K =p\sqrt{N} Q_p\,,
\eeq
and the angular momentum along the direction of rotation of each sphere is
\beq\label{Jodd}
J_I=\int_{\mc B_{p}} dV_{(p)}\mc K\; l_a\lp\frac{\partial}{\partial
\phi_I}\rp^a=V_{(p)}\mc K R_I\sqrt{\frac{p_I}{p}}\,,
\eeq
where now $V_{(p)}=\prod_I \Omega_{(p_I)}R_I^{p_I}$.
Using \eqref{MiKQ} and eliminating $\mc K$ we find
\beq
M-\sqrt{N}V_{(p)}Q_p=p\sqrt{N}V_{(p)}Q_p
=\lp\sum_{I=1}^\ell\frac{J_I^2}{R_I^2}\rp^{1/2}\,,
\eeq
and further eliminating the charge,
\beq
M=\frac{p+1}{p}\lp\sum_{I=1}^\ell\frac{J_I^2}{R_I^2}\rp^{1/2}\,,
\eeq
which generalizes \eqref{RJM}. Unless all radii are equal there is
no expression as simple as \eqref{QMJ}.

\bigskip

Tables \ref{II} and \ref{M} exhibit the limited number of allowed
possibilities in this restricted class of solutions in (uncompactified) type IIA/B string
theory and M-theory respectively. 

Note that, geometrically, the `large' odd-spheres
of the horizons are round even when higher-order corrections in the
blackfold construction are included. However, the `small' spheres
$s^{n+1}$ are round only in the leading test-brane approximation, and
will be distorted by the corrections from worldvolume curvature.

\begin{table}[t!]
\centering
\begin{tabular}{|c|c|c|}
\hline
Brane (IIA) & Worldvolume & $\perp$ Sphere   \\
\hline \hline
F1   &  $S^1$  & $s^7 $     \\ \hline
D2   & $\T^2$ & $s^6 $     \\ \hline
D4   & $S^3 \times S^1$ , $ \T^4$  & $s^4 $   \\ \hline
NS5   & $ S^5$  , $ S^3 \times \T^2$  & $s^3$    \\ \hline
D6   &  $ S^3 \times S^3$ , $ S^5\times S^1$  & $s^2$   \\ \hline
\end{tabular}
\hspace{0.8cm}
\begin{tabular}{|c|c|c|}
\hline
Brane (IIB) & Worldvolume & $\perp$ Sphere  \\
\hline \hline
D1   & $S^1$ & $s^7$   \\ \hline
F1   &  $S^1$ & $s^7 $     \\ \hline
D3   & $S^3$, $\T^3$ & $s^5$     \\ \hline
D5   & $S^5$  ,  $S^3 \times \T^2$ & $s^3 $   \\ \hline
NS5  & $S^5$ , $ S^3 \times \T^2 $ & $s^3 $\\ \hline
\end{tabular}
\bf\caption{\it A list of horizon topologies for stationary non-extremal
black holes in type IIA/IIB string theory based on the
singly-charged blackfolds of the theory with worldvolumes curved into
products of odd-spheres. The $s^{n+1}$ denotes the `small'
sphere in horizon directions orthogonal to the worldvolume. The number
$\ell$ of `large' odd-spheres spanned by the worldvolume is limited by \eqref{ellbound}.}
\label{II}
\end{table}

\begin{table}[t!]
\centering
\begin{tabular}{|c|c|c|c|}
\hline
Brane & Worldvolume  & $\perp$ Sphere \\
\hline \hline
M2   & $\T^2$ & $s^7 $     \\ \hline
M5   & $S^5$,   $ S^3 \times \T^2 $,  $\T^5$  & $s^4$    \\ \hline
\end{tabular}\bf
\caption{\it The analogue of Table \ref{II} in M-theory for M2 and M5 black branes.}
\label{M}
\bigskip
\end{table}

\paragraph{Solutions with Kaluza-Klein circles.} In the previous study the background
is globally flat Minkowski spacetime and
hence all the cycles that the blackfold wraps are contractible. We can
easily extend the analysis to situations where there is a number
$p_\circ$ of compact Kaluza-Klein circles, on which the brane is
supported by topology without requiring a centrifugal force to balance.
Compactifying on a torus $\sum_{A=1}^{p_\circ} dx^A dx^A$, for the velocity vector
we take
\beq\label{ukk}
u=\frac1{\sqrt{1-\tanh^2\bar\eta-v^2}}\lp\frac{\partial}{\partial t}+
\sum_{I=1}^\ell\Omega_I \frac{\partial}{\partial\phi_I}
+\sum_{A=1}^{p_\circ}v_A \frac{\partial}{\partial x^A}\rp\,,
\eeq
with 
\beq
\tanh^2\bar\eta=\sum_{I=1}^\ell\Omega_I^2 R_I^2\,,\qquad 
v^2=\sum_{A=1}^{p_\circ} v_A^2\,.
\eeq
The acceleration and extrinsic curvature vanish on the internal torus so the
extrinsic equations are non-trivial only in the non-compact directions.
The equilibrium conditions on odd-spheres are now
\beq
\Omega_I R_I=\sqrt{\frac{p_I}{p-p_\circ}}\tanh\bar\eta\,,
\eeq
\beq
\frac{\sinh^2\bar\eta}{1-v^2\cosh^2\bar\eta} =
\frac{p-p_\circ}{n}(1+ n N\sinh^2\alpha) \,.
\eeq
The velocity $v_A$ along the internal torus is arbitrary. When
$v^2=0$ we simply need to replace $p\to p-p_\circ$ in \eqref{RIOmIal}.

\section{Stability}\label{sec:GLstab}

The blackfold approach captures efficiently the potential stability or
instability of a black hole at long wavelengths. The general analysis is
complicated, but it simplifies for perturbations such that the intrinsic and extrinsic
equations decouple. We consider first the stability under intrinsic
perturbations whose wavelength is much smaller than the extrinsic length
scale $R$, so the curvature of the brane worldvolume can be neglected.

\subsection{Sound waves, Gregory-Laflamme and correlated stability}

Sound wave instabilities of the effective fluid, which correspond to
unstable fluctuations of the thickness of the black brane, have been
identified in ref.~\cite{Emparan:2009at} with the classical
Gregory-Laflamme instabilities \cite{Gregory:1994bj}. In addition,
ref.~\cite{Camps:2010br} has observed that the inclusion of the first
dissipative terms (associated to the contribution of shear and bulk
viscosities) provides an impressively good agreement with results
obtained by direct study of the Einstein equations, an agreement that
improves with increasing dimension $n$. 

Here we discuss how the addition of $p$-brane charge affects the
stability of a black $p$-brane. Since this charge does not add any local
degree of freedom to the fluid, it suffices to study the sound modes and
the analysis only differs from that of
\cite{Emparan:2009at,Camps:2010br} in the equation of state and
parameters of the fluid. In particular, ref.~\cite{Camps:2010br} finds
that for a generic relativistic fluid with bulk and shear viscosities
$\zeta$ and $\eta$, the linearized perturbations with wavenumber $k$ and
frequency $\omega =-i \Omega$
obey the
dispersion relation\footnote{In this section only, we use $\eta$,
$\zeta$, $\Omega$ and $k$ with a different meaning than in the rest of the article.}
\beq
\label{GLac}
\Omega=\sqrt{-c_s^2}k-\left[ \left(1-\frac{1}{p} \right)\frac{\eta}{s}+\frac{\zeta}{2s}\right] 
\frac{k^2}{\TT}+\OO(k^3)
~.
\eeq
$c_s$ is the speed of sound
\beq
\label{GLad}
c_s^2=\left(\frac{\d P}{\d \varepsilon} \right)_{Q_p}
\eeq
computed at fixed charge $Q_p$. An unstable mode exists when $c_s^2<0$.

\paragraph{Sound-mode instability.} To identify an instability we need
only focus on the first term, linear in $k$, in \eqref{GLac}
which arises at the perfect fluid level.  
Inserting the specific data of a charged dilatonic black $p$-brane solution \eqref{bffluid1} we find
\beq
\label{GLae}
c_s^2=-\frac{1}{n+1} \frac{1+(2-Nn)\sinh^2 \alpha}{1+\left(2-\frac{Nn}{n+1}\right)\sinh^2\alpha}
~.
\eeq
Since
\beq\label{salphi}
\sinh^2\alpha=\frac1N \frac{\Phi_p Q_p}{\mc T s}
\eeq
we can interpret the regimes of large or small $\alpha$ as regimes where
the charge-tensional component of the brane $\Phi_p Q_p$ is dominant, or
subdominant, relative to the thermal component $\mc T s$. Observe also
that given the upper bound on $N$, \eqref{Nbound}, and since $n,p\geq 1$,
we always have $Nn <2(n+1)$ so the denominator in \eqref{GLae} is always
positive. Thus stability requires that $(Nn-2)\sinh^2\alpha>1$, and we
find two different situations:
\begin{itemize}
\item[(i)] $Nn\leq 2$. In this case, $c_s^2<0$ always and there is no
regime of stability, no matter how close to extremality the brane is.
\item[(ii)] $2<Nn$. Stability requires
\beq
\label{GLaf}
\frac{1}{Nn-2}<\sinh^2 \alpha
~.
\eeq 
This is a regime where the energy of thermal
excitations is sufficiently smaller than the charge tension, which is
achieved as extremality is approached.
\end{itemize}

According to this, black $p$-branes in 
string theory ($N=1$, $D=10$) are always unstable to sound wave
perturbations for $p\geq 5$. When $p<5$, they become stable when they
are close enough to extremality that
\eqref{GLaf} holds.

The disappearance of the marginally unstable GL zero-mode when
\eqref{GLaf} is satisfied has been verified explicitly for several
classes of charged black branes in
\cite{Gregory:1994tw,Reall:2001ag,Hirayama:2002hn,Gubser:2002yi,Kang:2004hm,Miyamoto:2006nd,Miyamoto:2007mh}. 
However, it is important to realize that here we are not (yet) describing
these zero modes of finite wavelength, but rather identifying an
instability of hydrodynamic modes, with small frequency at very long
wavelengths.

\paragraph{Correlated stability.} The general thermodynamic relation
\beq
\label{GLak}
c_s^2=\left( \frac{\d P}{\d \varepsilon} 
\right)_{Q_p}=s\left( \frac{\d \TT}{\d \varepsilon} \right)_{Q_p}
=\frac{s}{c_{Q_p}}
\eeq
provides a direct relation between the speed of sound and the specific
heat at fixed charge $c_{Q_p}$. This is in accordance with the Correlated
Stability Conjecture \cite{Gubser:2000mm}. In more detail, the
blackfold formalism shows that local thermodynamical instabilities are
in one-to-one correspondence with long-wavelength, hydrodynamical
perturbations of black branes. However, the stability under modes that
do not have a hydrodynamic limit (\ie whose frequency does not vanish as
the wavelength goes to infinity) need not be correlated with local
thermodynamic stability.

\paragraph{Viscosity and the complete dispersion relation.}
We can now consider the effect of the term quadratic in $k$ in eq.~\eqref{GLac}.
This involves the ratios $\eta/s$ and $\zeta/s$ which
can be computed with a perturbative analysis of the Einstein equations
as in \cite{Camps:2010br}. Although we will not perform this analysis in this
paper, it is reasonable to anticipate that
\beq
\label{GLan}
\frac{\eta}{s}=\frac{1}{4\pi}\,, \qquad
\frac{\zeta}{s}=\frac{1}{2\pi} \left( \frac{1}{p} -c_s^2 \right)
\,.
\eeq
The first relation is the well-known universal value of $\eta/s$ for
event horizons in Einstein gravity theories. The second relation is a
similar universal relation
for the bulk viscosity proposed in \cite{Buchel:2007mf}.
This has been checked to hold in the near horizon limit of the black
branes we are considering \cite{Kanitscheider:2009as} and it may not be
unreasonable to expect that it remains valid when computed in the full
asymptotically flat geometry. Ref.~\cite{Camps:2010br} verified this
expectation for neutral black branes. 

We tentatively assume the validity of the equations
\eqref{GLan} and proceed to explore their implications. One reason why
this can be very interesting is that there are very few results for the
dispersion relation of GL modes of $p$-brane-charged black branes. Moreover the
conventional perturbation analyses have to be redone anew, numerically,
for each value of $n$, $N$, and the charge. In contrast, our methods give very
easily analytical results for all values of these parameters.

Inserting
\eqref{GLan} into the dispersion relation \eqref{GLac} we obtain
\beq
\label{GLao}
\Omega=\sqrt{-c_s^2}k-\frac{1-c_s^2}{4\pi} \frac{k^2}{\TT}+\OO(k^3)
~.
\eeq

When $c_s^2<0$, this equation is expected to be a quantitatively good
approximation to the Gregory-Laflamme dispersion relation of unstable
charged dilatonic branes at small wavenumbers, namely when $k/\TT\ll 1$.

For neutral black branes ($\alpha=0$) this extrapolation becomes a
better and better approximation to the exact result with increasing $n$
\cite{Camps:2010br} since for these branes $-c_s^2$ is small when $n\gg
1$. More generally, for charged black branes we may also expect that
whenever $-c_s^2$ is small, \ie close to the stability limit, the
dispersion relation is increasingly well
approximated by
\beq\label{GLexact}
\Omega\to \sqrt{-c_s^2}k-\frac{1}{4\pi}\frac{k^2}{\mc T}\,,
\eeq
since all unstable wavelengths are much longer than the thermal length.
Note, incidentally, that in contrast to neutral black branes, if we keep
$N$ fixed and make $n$ large this
does not make $c_s$ small. 

Numerical
results (but with low resolution) are available in
\cite{Gregory:1994bj} for the dispersion relation $\Omega(k)$ of five-dimensional
dilatonic strings, whose qualitative behavior
appears to agree well with our curves.

\paragraph{Threshold mode and critical behavior.}

The curve \eqref{GLexact} predicts the
appearance of a
Gregory-Laflamme threshold mode at
\beq
\label{GLap}
\frac{k}{\TT}\Big|_\mathrm{thr}=4\pi \sqrt{-c_s^2}
~.
\eeq

Numerical results for the threshold mode of
several black branes have been computed in 
\cite{Gregory:1994tw,Hirayama:2002hn,Gubser:2002yi,Kang:2004hm,Miyamoto:2006nd,Miyamoto:2007mh}. 
While we have not performed a detailed comparison, plots of
$k|_\mathrm{thr}$ as a function of the charge appear to show good
qualitative agreement with \eqref{GLap}. 

In the case of non-dilatonic branes,
refs.~\cite{Gubser:2002yi,Miyamoto:2006nd,Miyamoto:2007mh}
have observed that near the
critical value of the charge $Q_p^\mathrm{crit}$ at which $c_s^2=0$, the
threshold mode behaves like\footnote{The result is the same irrespective
of whether one fixes the mass, temperature, or $r_0$.}
\beq
k|_\mathrm{thr}\sim \lp Q_p-Q_p^\mathrm{crit}\rp^{0.5}\,,
\eeq	
\ie a power-law critical behavior with a numerically-determined
exponent close to the mean-field value $1/2$. Our
analytical result \eqref{GLap} yields exactly this exponent, in the
regime where we expect our approximations to be accurate. Furthermore, it
predicts that it applies also to dilatonic branes.

It would be interesting to have more numerical data to perform more
detailed quantitative comparisons of the threshold behavior and of the
complete dispersion curve.

\subsection{Elastic stability}

Transverse, elastic perturbations of the worldvolume geometry propagate
with speed
\beq
c_T^2=-\frac{P}{\vep}\,.
\eeq
For a charged dilatonic black $p$-brane solution  we find
\beq
\label{GLam}
c_T^2=\frac{1+nN \sinh^2\alpha}{1+n+nN \sinh^2 \alpha}
\eeq
which is always positive. In fact, as could be expected, the charge adds
to the brane tension and makes the brane more rigid, with the elastic
waves approaching lightspeed $c_T\to 1$ as extremality is approached.

\medskip

Thus we see that, as expected, the addition of $p$-brane charge that
cannot be redistributed along the worldvolume enhances the stability of
the black $p$-brane under perturbations that would make it
inhomogeneous. 

Although we have not performed a complete analysis of
coupled intrinsic-extrinsic perturbations, it seems reasonable to expect
that generic $p$-brane-charged blackfolds with $Nn\leq 2$, when they are
near-extremal in the sense that $\sinh^2\alpha\gg 1/(Nn-2)$,
correspond to stable black holes\footnote{Intrinsic modes that are close
to being marginally stable might in principle become unstable through
coupling to extrinsic perturbations. This is why we require to be well
inside the intrinsic stability regime.}.

\section{Blackfolds with brane currents}\label{sec:qcurrents}

D-branes in string theory can support worldvolume currents that
correspond to charges of strings or other D-branes `dissolved' in their
worldvolume. In the DBI description, they are captured as classical solutions of
the Born-Infeld worldvolume gauge fields, \ie as open-string
excitations. In the blackfold approach, instead, they appear as
closed-string modes, in very much the same manner as in the AdS/CFT
correspondence. The local gauge symmetry on the worldvolume is not
present, and the black brane supports a global conserved current that
sources the background spacetime gauge field. The effective theory of blackfolds
allows to describe thermal excitations of the worldvolume of the D-brane
with these currents, something that the classical DBI action cannot do.
Moreover, in the same manner as in sec.~\ref{subsubsec:momwaves}, the
blackfold method can also describe some extremal configurations that lie
outside the reach of the DBI action.

Such brane currents greatly expand the possible dynamics of blackfolds
and the new classes of black holes that can be constructed out of them.
In particular, on a spatially compact worldvolume a $q$-brane current
with $q\geq 1$ gives rise to a dipole, while a 0-brane (\ie a particle)
current yields a charge.

In this section we consider generic black $p$-branes that carry
$q$-brane currents on their worldvolume.
At the perfect fluid level much of the analysis can be done for a
multi-charge brane with several $q$-brane currents. At a later point we
shall focus mostly on configurations where in addition to the $p$-brane
charge there is also $q=0$ or $q=1$ brane charge.

\subsection{Blackfold fluids with $q$-brane currents}

We consider the dynamics of a fluid in a worldvolume $\mc
W_{p+1}$ that supports one or several conserved $q$-brane currents. In
this paper we only sketch the main properties of these theories --- many
details are similar to \cite{Caldarelli:2010xz}. Furthermore, we confine
ourselves from the outset to the perfect fluid approximation. This is
enough for the purposes of constructing stationary configurations.

Each $q$-brane current foliates $\mc
W_{p+1}$ into sub-worldvolumes $\mc C_{q+1}\subset \mc W_{p+1}$. On each of
these we consider a unit $(q+1)$-form $\hat V_{(q+1)}$ so that the current is
\beq
J_{(q+1)}=\mc Q_{q}\hat V_{(q+1)}
\eeq
with $\mc Q_{q}$ being the $q$-brane density.
The conservation equation
$d\ast J_{(q+1)}=0$ requires that $\ast \hat V_{(q+1)}\wedge 
d\ast\hat V_{(q+1)}=0$, which is ensured when the currents are
hypersurface-forming, as we will assume in what follows.

The currents make the perfect fluid anisotropic, and in particular they
induce differences between the pressures in directions parallel and
transverse to them.\footnote{In the context of AdS/CFT such systems have
been studied recently in \cite{Mateos:2011tv}.} These
differences are due to the effective tension $\Phi_q \mc Q_q$ along the
current, where $\Phi_q$ is
the chemical potential conjugate to the brane density
$\mc Q_q$, so the thermodynamic relations 
\beq
d\vep=\mc T d s+{\sum_q}' \Phi_q d\mc Q_q \,, 
\eeq 
are satisfied locally in the fluid, and the stress-energy tensor can be
written as\footnote{While we do not have a complete proof that this is
the universal form for any charged brane, we believe it is generic for
large classes of branes and it definitely applies to the black branes of
IIB/IIA/11D supergravities and their toroidal compactifications.}
\beq\label{tabgen}
T_{ab}=\mc T s\, u_a u_b -\mc G\,\gamma_{ab} -\sum_q \Phi_q \mc Q_q h^{(q)}_{ab}\,.
\eeq
The prime in ${\sum_q}'$ indicates that $q=p$ (the case considered in
sec.~\ref{sec:1charge}) is excluded from the sum,
and is instead included in $\sum_q$.
In eq.~\eqref{tabgen} we have introduced 
\beq
\mc G=\vep -\mc T s -\sum_q \Phi_q \mc Q_q
\eeq
as the local Gibbs free energy density and $h^{(q)}_{ab}$ as the
projector (induced metric) onto
$\mc C_{q+1}$. For instance, $h^{(0)}_{ab}=-u_a u_b$, and
$h^{(p)}_{ab}=\gamma_{ab}$.

Note that: (i) the brane densities $\mc Q_q(\si)$
are only `quasi-local' in
the sense that they are constant along $\mc
C_{q+1}$ and can vary only in the $p-q$ directions transverse to the current. In
particular, when
$q=p$ the charge
$\mc Q_p=Q_{p}$ is a non-dynamical global constant and all the remarks
to this effect made in Sec.~\ref{sec:1charge} apply. (ii) The brane currents need
not be `nested' in the
sense that we need not have $\mc
C_{q+1}\subseteq \mc C_{q'+1}$ for two currents with $q\leq q'$. For instance we may have two string
currents each along a different direction.

A general relation valid for the black branes of IIA/IIB/11D
supergravities and their toroidal compactifications,
is 
\beq\label{localsmarr}
\vep=\frac{n+1}{n}\mc T s+\sum_q \Phi_q \mc Q_q\,.
\eeq
This is the conventional Smarr relation for the
$(n+3)$-dimensional static black hole that results if we compactify
the black $p$-brane on a $p$-torus, which has as many charges as
$q$-brane currents.
An expression equivalent to \eqref{localsmarr} is
\beq\label{GTs}
\mc G=\frac{1}{n}\mc T s\,.
\eeq
Thus the general form of the stress-energy tensor of charged blackfold fluids is
\beq\label{gentabq}
T_{ab}=\mc T s\lp u_a u_b -\frac{1}{n}\,\gamma_{ab}\rp -\sum_q \Phi_q
\mc Q_q h^{(q)}_{ab}\,.
\eeq

The extrinsic equations
for the $p$-brane, $K_{ab}{}^\rho T^{ab}=0$, now take the form
\beq\label{exteqqbrane}
\mc T s\perp^\rho{}_\mu\dot u^\mu = \frac{1}{n}\mc T s K^\rho+
\perp^\rho{}_\mu\sum_q \Phi_q \mc Q_q K^\mu_{(q)}
\eeq
where 
\beq
K^\mu_{(q)}=h^{ab}_{(q)}{K_{ab}}^\mu
\eeq 
is the mean curvature vector of the embedding of
$\mc C_{q+1}$ in the background spacetime.

Locally, the velocity $u^a$ corresponds to a boost along the $p$-brane.
If this boost is along directions orthogonal to the $q$-brane current,
then $\hat V_{(q+1)}$ and $h_{ab}^{(q)}$ transform non-trivially under
the boost. If the boost is parallel to the current, then they are left
invariant by it.

\paragraph{Charges, potentials, and thermodynamics.} The intrinsic
equations are the equations of continuity of charge and energy-momentum
on $\mc W_{p+1}$. Instead of repeating their analysis, we shall jump
directly to the solution for stationary configurations. These have the
velocity $u$ aligned with a Killing vector $k$, and their temperature
redshifts locally according to \eqref{Tshift}. For these solutions the
potential $\Phi_q(\si)$ does not depend on the Killing time coordinate
and one can also show that it can not depend either on directions
transverse to the current \cite{Caldarelli:2010xz} --- this is the familiar equipotential
condition for electric equilibrium.
Thus, the integration over spatial directions
along the current gives a constant
\beq\label{globPhi}
\Phi_H^{(q)}=\int d^q\si |h^{(q)}(\sigma^a)|^{1/2}\Phi_{q}(\si)
\eeq
that corresponds to the global $q$-brane
potential for the stationary configuration. This
integration over the spatial directions of the worldvolume of the
$q$-brane is of the same kind as in \eqref{Phidef}.

As before, we assume that $k$ is of the form \eqref{kxichi}, where
$\xi$ foliates $\mc W_{p+1}$ into spatial sections
$\mc B_p$ with unit
timelike normal $n^a$, \eqref{nxi}.
The total mass, angular momenta and entropy are obtained as in
\eqref{MJS}. 

In order to obtain the $q$-brane charge we need to
consider the spatial sections of the $q$-brane worldvolume $\mc C_{q+1}$
that are orthogonal to $n^a$. On these, the
unit $q$-form $\omega_{(q)}$ orthogonal to $n^a$ is
\beq
\omega_{(q)}=\frac{-\hat V_{(q+1)}\cdot n}{\sqrt{-h^{(q)}_{ab}n^a n^b}}\,.
\eeq
The total
$q$-brane charge $Q_q$ is obtained by integrating its density over
the directions transverse to the $q$-brane current,
\beq
\label{Qq}
Q_q=
-\int_{\mc B_{p-q}}dV_{(p-q)}J_{(q+1)}\cdot (n\wedge\omega_{(q)})
=\int_{\mc B_{p-q}}dV_{(p-q)} \sqrt{-h^{(q)}_{ab}n^a n^b}\;\mc Q_q(\si)\,.
\eeq
The global potential \eqref{globPhi} is\footnote{These expressions for
$Q_q$ and $\Phi_H^{(q)}$
generalize and simplify the ones given in \cite{Caldarelli:2010xz} for $q=0,1$.}
\beq
\label{PhiHq}
\Phi_H^{(q)}=\int d V_{(q)}\frac{R_0}{\sqrt{-h^{(q)}_{ab}n^a n^b}}\Phi_q(\si)\,.
\eeq
If the velocity is aligned with the $q$-brane then we can always write
$\hat V_{(q+1)}=n\wedge \omega_{(q)}$ and $\sqrt{-h^{(q)}_{ab}n^a
n^b}=1$, and the potential only undergoes a gravitational redshift
$R_0$. Otherwise \eqref{PhiHq} includes
also a Lorentz-boost redshift factor. For instance, when the boost is
orthogonal to the current, $\sqrt{-h^{(q)}_{ab}n^a n^b}=-u^a n_a$.

By a straightforward extension of previous results, it is possible to obtain
the extrinsic equations \eqref{exteqqbrane} for
stationary fluid configurations on the brane from the variation of the action
\beqa
I&=&-\int_{\mc
W_{p+1}}d^{p+1}\si\sqrt{-\gamma}\;\mc G\nn\\
&=&-\Delta t\lp M-TS-\Omega J-\sum_q\Phi_H^{(q)}Q_q\rp
\eeqa
keeping $T$, $\Omega$, $\Phi_H^{(q)}$
constant.  
Then the global first law
\beq
dM=TdS+\Omega dJ +\sum_q\Phi_H^{(q)}dQ_q
\eeq
for variations of the brane embedding is equivalent to these extrinsic
equations.

When \eqref{GTs} is integrated over the blackfold worldvolume, it implies \eqref{acts}.
Using this we can derive the integrated expression
\beq\label{presmarr2}
(D-3)M-(D-2)(TS+\Omega J)-\sum_q(D-3-q)\Phi_H^{(q)} Q_q
=\mbT\,,
\eeq
which, in Minkowski backgrounds, reduces to the Smarr relation for
asymptotically flat black
holes when $\mbT=0$. Thus the latter must follow as a consequence
of the extrinsic equations.

\paragraph{Intrinsic solution for $q=0,1$ currents.} 

In the previous analysis each current enters independently. 
Moreover, the presence of $p$-brane charge does not
introduce any new degree of freedom. Therefore, when the brane
has any 0-branes or strings in the worldvolume the general analysis of stationary
solutions with these charges in \cite{Caldarelli:2010xz} carries over directly.

When $q=0$ the complete solution to the intrinsic equations is
obtained by taking the 
0-brane potential to redshift like the temperature,
\beq
\Phi_{0}(\si)=\frac{\Phi_{H}^{(0)}}{|k|}\,,
\eeq
with constant $\Phi_{H}^{(0)}$. 

For string currents, $q=1$, an intrinsic solution in
which the geometry along the currents, $h^{(1)}_{ab}$, is fully specified, can
be given explicitly when
the strings lie along a spatial Killing vector $\psi$ that commutes with
$k$. Then, introducing
\beq\label{zetapsi}
\zeta^a=\psi^a+(\psi^b u_b)u^a
\eeq
we have
\beq\label{h1K1}
h^{(1)}_{ab}=-u_a u_b+\frac{\zeta_a \zeta_b}{|\zeta|^2}\,,\qquad 
|h^{(1)}|^{1/2}=|k||\zeta|\,,
\qquad K_{(1)}^\rho=
-\lp g^{\rho\mu}- h_{(1)}^{\rho\mu}\rp\partial_\mu\ln(|k||\zeta|)\,.
\eeq
Assuming that $\psi$ (and hence $\zeta$) 
has compact orbits of periodicity
$2\pi$, the potential is
\beq
\Phi_{1}(\si)=\frac{\Phi^{(1)}_H}{2\pi|h^{(1)}(\sigma^a)|^{1/2}}\,,
\eeq
with constant $\Phi_{H}^{(1)}$.

\subsection{Extremal limits} 
\label{subsec:extq} 

\paragraph{Extremal branes with subluminal worldvolume velocity.} At extremality we have $\mc T s\to 0$. The
conventional limit in which the velocity is not scaled to light speed,
takes \eqref{gentabq} into the form
\beq\label{extnnTab}
T_{ab}= -\sum_q \Phi_q \mc Q_q h^{(q)}_{ab}\,.
\eeq
In contrast to the case in sec.~\ref{subsubsec:dirac} with only
$p$-brane charge, if the velocity
lies along a direction orthogonal to at least one of the currents, it
is a physical mode that cannot be gauged away. The velocity remains
subluminal in this limit.

Assuming that at stationarity $u^a$ lies along a Killing direction, the
extremal limit
$TS\to 0$ of the Smarr relation (with $\mbT=0$) and
the vanishing of the extremal action $I$ allow to derive
\beq\label{MQOmJ}
M=\sum_q (q+1)\Phi_H^{(q)} Q_q\,,\qquad \Omega J=\sum_q q\Phi_H^{(q)} Q_q\,.
\eeq
These generalize expressions found in \cite{Caldarelli:2010xz}. When
there is only a 0-brane current, as considered in
\cite{Caldarelli:2010xz}, the kinetic energy term $\Omega J$ must
vanish. However we see that, in combination with higher-brane currents,
0-brane-charged extremal blackfolds need not be static, in fact the
rotation is required in order to balance the tension that these higher
currents create.

In configurations of equilibrium in Minkowski backgrounds, the total
rapidity $\eta$ is determined by the condition $\mbT=0$. If we
denote by $q_\perp$ and
$q_\|$ those currents that are orthogonal and parallel to the
velocity, we have
\beq
(\gamma^{ab}+n^a n^b)h^{(q_\perp)}_{ab}=-\sinh^2\eta +q_\perp\,,\qquad
(\gamma^{ab}+n^a n^b)h^{(q_\|)}_{ab}=q_\|\,.
\eeq
Then $\mbT=0$ can easily be seen to imply that
\beq
\int_{\mc B_p}dV_{(p)}\sinh^2\eta\sum_{q_\perp}\left(\Phi_{q_\perp}\mc Q_{q_\perp}\right)
=\sum_q q \Phi_H^{(q)}Q_q\,.
\eeq
In cases for which the rapidity $\eta$ is constant on $\mc B_p$ this equation fixes
its value at equilibrium to be
\beq\label{extsheta}
\sinh^2\eta=\frac{\sum_q q \Phi_H^{(q)}Q_q}{\sum_{q_\perp}\Phi_H^{(q_\perp)}Q_{q_\perp}}\,.
\eeq

\paragraph{Null-wave branes.}
The limit where the fluid velocity becomes lightlike 
requires that we keep
\beq
\sqrt{\mc T s}\;u^a=\sqrt{{\mc K}}\; l^a
\eeq
finite, with $l^a$ approaching a null vector and ${\mc K}$ representing a
momentum density. Then
\beq
T_{ab} = {\mc K} l_a l_b -\sum_q \Phi_q \mc Q_q h^{(q)}_{ab}\,.
\eeq
When $q=0$, for which $h^{(0)}_{ab}=-u_a u_b$, this limit requires that
we also send $\Phi_0 \mc Q_0\to 0$ so the 0-brane charge disappears in
the limit. More generally, whenever the boost
occurs along directions orthogonal to the $q$-brane the metric $h^{(q)}_{ab}$
degenerates and the $q$-brane charge must disappear in the limit. On
the other hand, if the boost occurs along the $q$-brane, the metric
$h^{(q)}_{ab}$ remains invariant and non-degenerate, and the $q$-brane charge
survives the limit. Thus, extremal branes with a null-wave only support
currents parallel to the direction of the wave.

Again, we can use the condition of absence of tensional energy to obtain
general expressions for equilibrium configurations in Minkowski
backgrounds, which generalize those found in
sec.~\ref{subsubsec:momwaves}. Normalizing $n^al_a=-1$ we find
\beq\label{momcharge}
\int_{\mc B_p}dV_{(p)}\mc K=\sum_q q \Phi_H^{(q)}Q_q\,.
\eeq
The mass is
\beq\label{masscharge}
M=\int_{\mc B_p}dV_{(p)}\mc K+\sum_q \Phi_H^{(q)}Q_q=\sum_q(q+1)\Phi_H^{(q)}Q_q\,.
\eeq
These are analogous to \eqref{MQOmJ}. When $\chi^a l_a=\mc R$ is
constant on $\mc B_p$ we obtain a generalization of \eqref{MQJ},
\beq
\frac{J}{\mc R}=\sum_q q \Phi_H^{(q)}Q_q\,.
\eeq

For branes such that at extremality $\Phi_q\to 1$, which we believe
occurs when they are marginally bound (like in the D0-D4 and D1-D5
systems below), these expressions apply in the form 
\beq\label{MVQ}
M=\sum_q (q+1)V_{(q)} Q_q\,,\qquad \frac{J}{\mc R}=\sum_q qV_{(q)} Q_q
\eeq
with $V_{(q)}$ the spatial volume along the $q$-brane.

\paragraph{Extremality vs.\ supersymmetry.} Although expressions such as
\eqref{MQOmJ} and \eqref{MVQ} may be reminiscent of BPS relations, one
must bear in mind that we are considering that $\mc B_p$ is compact and
hence, in Minkowski space, contractible to a point. Therefore $Q_q$ for
$q\geq 1$ correspond to dipoles and not to conserved charges, which are
the ones entering the BPS bounds. Generically, these configurations are
not supersymmetric.

The charge of $q=0$ currents is a net, conserved charge. Then, if there
are only 0-brane currents (excluding also a $p$-brane current), the first
relation in \eqref{MVQ} is indeed the BPS relation. As discussed in
\cite{Caldarelli:2010xz}, the resulting (singular, static) blackfolds are supersymmetric.

\subsection{Products of odd-spheres}
\label{subsec:oddq}

The most straightforward solutions to construct are those in which the
blackfold wraps a product of odd-spheres, like we did in
sec.~\ref{sec:Dpoddsphere}. The general analysis, although
essentially straightforward, can become very cumbersome, so we shall confine
ourselves to a few simple and interesting instances. Many other
particular cases can be readily worked out.

\paragraph{$(0$-$p)$-brane solutions.} We begin with the case where we
have both 0- and $p$-brane currents. The velocity field and the
extrinsic curvatures for the product of odd-spheres are like in \eqref{uoddsph}, \eqref{Kri}. For
the 0-brane (a particle) the extrinsic curvature vector is the acceleration
$K^\rho_{(0)}=-{\perp^\rho}_\mu \dot u^\mu$, and we find that the
extrinsic equations \eqref{exteqqbrane} are solved
by 
\beq\label{OmIRI}
\Omega_I R_I=\sqrt{\frac{p_I}{p}}\tanh\eta\,,
\eeq
\beq\label{shetasolqpnext}
\sinh^2\eta =\frac{p}{n}\frac{\mc T s +n\Phi_p Q_p}{\mc T s+\Phi_0 
\mc Q_0}\,.
\eeq
This reproduces \eqref{soleta} when $\mc Q_0=0$, and the result in
\cite{Caldarelli:2010xz} when $Q_p=0$. In the extremal limit in which $\mc T s=0$ we
find
\beq\label{shetasolqp}
\sinh^2\eta =\frac{p\Phi_p Q_p}{\Phi_0 
\mc Q_0}\,,
\eeq
which is what \eqref{extsheta} gives when the blackfold is homogeneous
and hence $\Phi_H^{(q)}Q_q=V_{(p)}\Phi_q \mc Q_q$.

\paragraph{$(1$-$p)$-brane solutions.} In this case the string current
introduces an anisotropy on the worldvolume, and there are different
configurations depending on how the strings are aligned relative to the
velocity field. Taking the string to lie along any spatial Killing vector, all
these possibilities can be obtained using the formalism of
eqs.~\eqref{zetapsi} and \eqref{h1K1}. 

For definiteness and simplicity, we consider the configuration in
which the string current is parallel to the velocity field \eqref{uoddsph}.
Hence we choose $\psi=\sum_I \Omega_I \partial_{\phi_I}$ in
\eqref{zetapsi}, and obtain
\beq
\zeta=\sinh^2\eta\frac{\partial}{\partial t}
+\cosh^2\eta\sum_I \Omega_I \frac{\partial}{\partial \phi_I}
\eeq
with $|\zeta|=\sinh\eta$ and $|h^{(1)}|^{1/2}=\tanh\eta$. The extrinsic
equations are now solved by
\beq\label{1pextsol}
\Omega_I R_I=\sqrt{\frac{p_I}{p}}\tanh\eta\,,\qquad
\sinh^2\eta =\frac{p}{n}+\frac{\Phi_1 \mc Q_1+p\Phi_p Q_p}{\mc T s}
\,.
\eeq
Again, the results in \cite{Caldarelli:2010xz} for $Q_p=0$ are reproduced. 

We can describe these configurations as having helical strings,
extending along the orbits of $\psi$, which are smeared on the
worldvolume of the $p$-brane. When $p=1$ we can indeed obtain the
generalization of the helical strings in \cite{Emparan:2009vd} that
carry string charge, and the equilibrium condition for them
is\footnote{For simplicity we are not distinguishing if there is one or
two types of strings.}
\beq
\sinh^2\eta=\frac{1}{n}+\frac{\Phi_1 \mc Q_1}{\mc T s}\,.
\eeq
As shown in \cite{Emparan:2009vd}, these helical strings generically break the Cartan
subgroup of the $D$-dimensional rotation group, possibly down to a
single $U(1)$. However, when $p>1$ the smearing of the strings over the
worldvolume of the $p$-brane (which itself is not helical) restores the
entire Cartan subgroup, which then remains unbroken.

Since the strings are parallel to the fluid velocity, in the extremal
limit $\mc T s\to 0$ this velocity becomes null $\eta\to\infty$, with
the momentum $\mc K=\mc T s\sinh\eta\cosh\eta$ remaining finite.
In this limit eqs.~\eqref{1pextsol} become
\beq\label{locK}
\mc K =\Phi_1 \mc Q_1+p\Phi_p Q_p
\eeq
which, when integrated (trivially) over the worldvolume reproduces
\eqref{momcharge}. The
component of the velocity on the $I$-th sphere is given by
\eqref{OmRIext}, the null vector is \eqref{lodd}, and 
the angular momentum along the direction of rotation of each sphere is \eqref{Jodd}.
Together with \eqref{masscharge}, we obtain
\beq
M-\Phi_H^{(1)}Q_1-\Phi_H^{(p)}Q_p=\Phi_H^{(1)}Q_1+p\Phi_H^{(p)}Q_p
=\lp\sum_I\frac{J_I^2}{R_I^2}\rp^{1/2}\,.
\eeq

\medskip

It is also possible to construct more general configurations in which
the string current is not aligned
with the
velocity, \ie $\psi=\sum_i \varsigma_I\partial_{\phi_I}$ with constant
coefficients $\varsigma_I$.
However,
even in simple cases the expressions one obtains are not too illuminating,
so we omit their study (related examples were constructed in
\cite{Caldarelli:2010xz}).

\paragraph{Solutions with Kaluza-Klein circles.} 
In a flat background with a compact torus $\T^{p_\circ}$, we can wrap the
brane with velocity vector \eqref{ukk}.

For $(0$-$p)$-brane systems, the equilibrium condition on odd-spheres is then
\beq\label{OmIRI2}
\Omega_I R_I=\sqrt{\frac{p_I}{p-p_\circ}}\tanh\bar\eta\,,
\eeq
\beq
\frac{\sinh^2\bar\eta}{1-v^2\cosh^2\bar\eta} =
\frac{p-p_\circ}{n}\frac{\mc T s +n\Phi_p Q_p}{\mc T s+\Phi_0 \mc Q_0}\,.
\eeq

For $(1$-$p)$-brane systems with velocity along the 1-brane, so that
$\psi=\sum_I \Omega_I \partial_{\phi_I}+\sum_A v_a\partial_A$, at
equilibrium we find eq.~\eqref{OmIRI2} and $\tanh\bar\eta$ determined
by solving
\beq
\frac{\sinh^2\bar\eta}{1-v^2\cosh^2\bar\eta} =\frac{p-p_\circ}{n}
+\frac1{\mc T s}\lp \frac{\Phi_1 \mc Q_1}{1+v^2/\tanh^2\bar\eta}
+(p-p_\circ) \Phi_p Q_p\rp\,.
\eeq
Note that, in our construction, $v\neq 0$ implies that the 1-brane wraps
the internal torus as well as the spheres in the non-compact space.

When
$v^2=0$ we simply need to replace $p\to p-p_\circ$ in \eqref{OmIRI},
\eqref{shetasolqpnext} and \eqref{1pextsol}.

\section{D0-D$p$ and F1-D$p$ blackfolds \label{sec:twocharge} }

Now we turn to some explicit examples of blackfold fluids with
lower-brane currents supported on their worldvolume. For concreteness we
focus first on two simple cases based on two-charge branes in
ten-dimensional type IIA/IIB string theory, namely blackfolds based on
the D0-D$p$ ($p=2,4,6$) and F1-D$p$ ($p \geq1 )$ brane solutions
respectively. These correspond to charged $p$-branes with
0-brane (particle) charge current or 1-brane (string) charge current on
their worldvolumes. 

\subsection{Effective fluids and stability of charge waves \label{sec:2chargefluid}}

We start by providing the local thermodynamics of the effective
blackfold fluids for D0-D$p$ and F1-D$p$ branes in ten-dimensional type
II string theory (we do not consider smeared D$p$-brane charges here,
but the considerations are easily generalized to that case). The local
thermodynamic quantities are obtained from the corresponding type II
supergravity solution (\ie the metric and gauge potential) of the
corresponding two-charge black brane solutions. These are given in
appendix~\ref{app:branesolns}.

While all these fluids satisfy the general relations \eqref{localsmarr}
and \eqref{GTs}, they differ in specific properties of their equations
of state. The most salient difference is visible in the sign of the
isothermal permittivity of the local charge (D0 or F1)
\beq
\epsilon_q\equiv\lp\frac{\partial \Phi_q}{\partial \mc Q_q}\rp_{Q_{p},\mc T}=
\lp\frac{\partial^2 \mc F}{\partial \mc Q_q^2}\rp_{Q_{p},\mc T}
\eeq
where $\mc F=\vep -\mc T s$ is the Helmholtz free energy density. This
permittivity measures the stability against fluctuations 
of the $q$-brane density in the fluid (which is a different kind of perturbation than
the sound modes associated to the Gregory-Laflamme instability), and is
directly related to whether the $q$-branes form actual bound states with
the $p$-branes, or instead are only marginally or unstably bound. We
will compute this quantity for each of the fluids discussed below for
the general non-extremal brane system (see also Appendix \ref{app:iso}
for details) and comment in connection with this in the extremal limit.
Also note that for all the effective fluids considered here, the
relation \eqref{Gbrane} for the Gibbs free energy density continues to hold, as
seen using \eqref{GTs}. 

\subsubsection*{D0-D2 fluid}

The basic thermodynamic quantities are
\begin{subequations}
\label{bs2af}
\beq
\vep=\frac{\Omega_{(6)}}{16\pi G} r_0^5 (6+5 \sinh^2\al)\,,
\eeq
\beq
\mc T=\frac{5}{4\pi r_0\cosh\al}\,,\qquad
s=\frac{\Omega_{(6)}}{4G} r_0^{6} \cosh\al\,,
\eeq
\beq
Q_{{\rm D}2}=\frac{5 \Omega_{(6)}}{16\pi G} r_0^5 \sinh\al \cosh\al\cos\theta\,,
\qquad \mc Q_{{\rm D}0}=Q_{{\rm D}2}\tan\theta\,,
\eeq
\beq
\Phi_{{\rm D}2}=\tanh\al \cos\theta\,, \qquad \Phi_{{\rm D}0}=\Phi_{{\rm D}2} \tan\theta\,.
\eeq
\end{subequations}
The extremal limit takes $r_0\to 0$, $\alpha\to\infty$ keeping
$r_0^5\sinh^2\alpha$ finite. In this limit the entropy vanishes
and $\vep^2=\QQ_{\rm D0}^2+Q_{\rm D2}^2$.

The permittivity of D0 charge density (for fixed D2 charge) is
\beq
\label{permiaa}
\epsilon_{\rm D0}=\sqrt{\frac{-7+12x+\sqrt{49-24x}}{-7+2x+\sqrt{49-24x}}}
\frac{1}{(\QQ_{\rm D0}^2+Q_{\rm D2}^2)^\frac32}
\left( Q_{\rm D2}^2+\QQ_{\rm D0}^2
\frac{10x}{46x-21 +3 \sqrt{49-24x}} \right)\,.
\eeq
Here we have defined the non-extremality parameter
\beq
\label{permiab}
x \equiv 1-\frac{\QQ_{\rm D0}^2+Q_{\rm D2}^2}{\varepsilon^2}
~.
\eeq
which by definition lies in the range $x\in [0,1)$. At extremality (with zero momentum), namely when $x \to 0$,
the permittivity \eqref{permiaa} bcomes 
\beq
\epsilon_\mathrm{D0}
=\frac{Q_{{\rm D}2}^2}{\lp Q_{{\rm D}2}^2+\mc Q_\mathrm{D0}^2\rp^{3/2}}>0
\eeq
This indicates that the system is stable to fluctuations of the charge:
it tends to distribute uniformly on the worldvolume, which is a
consequence of the fact that the D0-D2 form a true (1/2 BPS) bound state. Thus the D0-D2
brane can support stable waves of D0 charge in its worldvolume. Their
velocity is determined by this permittivity.

Beyond the extremal limit the D0 charge permittivity remains positive. As a function of the
non-extremality parameter $x$ it is positive and monotonic when we vary $x$
at fixed $\QQ_{\rm D0}, Q_{\rm D2}$ and diverges at $x=1$.

\subsubsection*{D0-D4 fluid}

The basic thermodynamic quantities are
\begin{equation}
\varepsilon = \frac{\Omega_{(4)}}{16 \pi G} r_0^3 (4+3 \sinh^2 \alpha_0+3\sinh^2 \alpha_4)
\end{equation}
\begin{equation}
\CT = \frac{3}{4\pi r_0 \cosh \alpha_0 \cosh \alpha_4} \spa s = \frac{\Omega_{(4)}}{4G} r_0^{4} \cosh \alpha_0 \cosh \alpha_4
\end{equation}
\begin{equation}
\mc Q_{{\rm D}i} = \frac{\Omega_{(4)}}{16 \pi G} 3 r_0^{3} \cosh \alpha_i \sinh \alpha_i
\spa \Phi_{{\rm D}i} =  \tanh \alpha_i
\end{equation}
with $i=0,4$. 

In the extremal limit, $r_0\to 0$, $\alpha_i\to\infty$, with
$r_0^3\sinh^2\alpha_i$ finite. In this limit the entropy vanishes
and $\vep=\QQ_{\rm D0}+Q_{\rm D4}$.

The permittivity of D0 charge is
\beq
\label{permiba}
\epsilon_\mathrm{D0}=\frac{1}{\QQ_\mathrm{D0}}\frac{\tanh \alpha_0 \left(\cosh^2 \alpha_4-2\right)}
{2-\cosh^2 \alpha_0+\sinh^2 \alpha_4(-8\cosh^2 \alpha_0+7)}
~.
\eeq
At extremality this permittivity becomes 
\beq
\epsilon_\mathrm{D0}=0
\eeq
which indicates that the system is marginally stable to fluctuations of
the charge. This is a consequence of the fact that the D0-D4 form a (1/4  BPS) marginal bound state.
Beyond extremality, say as we vary $\alpha_0$ at fixed $\QQ_\mathrm{D_0},
Q_\mathrm{D4}$, we find that $\epsilon_\mathrm{D0}<0$ except in a finite range of $\alpha_0$ where
it is positive (see App.~\ref{app:iso}). 

\subsubsection*{D0-D6 fluid}

In this case the basic thermodynamic quantities are
\begin{equation}
\varepsilon = \frac{\Omega_{(2)}}{16 \pi G} r_0 ( 2 +  \sinh^2 \alpha_0 +  \sinh^2 \alpha_6 )
\end{equation}
\begin{equation}
\CT = \frac{1}{4\pi r_0 \cosh \alpha_0 \cosh \alpha_6}
\frac {2(\cosh^2 \alpha_0 + \cosh^2 \alpha_6)} { (\cosh^2 \alpha_0 +1) (\cosh^2 \alpha_6 +1)}
\end{equation}
\begin{equation}
s =
\frac{\Omega_{(2)}}{4G} r_0^{2} \cosh \alpha_0 \cosh \alpha_6  \frac{ (\cosh^2 \alpha_0 +1) (\cosh^2 
\alpha_6 +1)} {2(\cosh^2 \alpha_0 + \cosh^2 \alpha_6)}
\end{equation}
\begin{equation}
\mc Q_{{\rm D}i} 
=  \frac{\Omega_{(2)}}{16 \pi G} r_0 \cosh \alpha_i \sinh \alpha_i \sqrt{ \frac{ \cosh^2 \alpha_i +1}
{\cosh^2 \alpha_0 + \cosh^2 \alpha_6} }
\end{equation}
\begin{equation}
\Phi_{{\rm D}i} =  \tanh \alpha_i
\sqrt{ \frac{\cosh^2 \alpha_0 + \cosh^2 \alpha_6} {\cosh^2 \alpha_i +1} }
\end{equation}
with $i=0,6$. 

Among all D0-D$p$ systems, this one is unique in having finite entropy density
in the extremal limit, in which $r_0\to 0$, $\alpha_i\to \infty$, with
$r_0\sinh^2\alpha_i$ finite, so
\beq
s=16\pi G\mc Q_\mathrm{D0}Q_\mathrm{D6}
\eeq
and $\vep^{2/3}=Q_\mathrm{D6}^{2/3}+\mc Q_\mathrm{D0}^{2/3}$.

The expression for the D0 charge permittivity is now more complicated and is given in eq.~\eqref
{permica}. At extremality the permittivity is
\beq
\label{permicd}
\epsilon_\mathrm{D0}=
-\frac{1}{2}\frac{\mc Q_\mathrm{D0}^{-4/3} Q_\mathrm{D6}^{2/3}}
{\sqrt{Q_\mathrm{D6}^{2/3}+\mc Q_\mathrm{D0}^{2/3}}}<0
\eeq
which indicates that the system is unstable to fluctuations of the charge:
it is favorable for the charge to clump. This is a
consequence of the fact that the D0-D6 repel each other: the D0 charge
tends to clump so that the D0-branes can be expelled. Further analysis of the expression \eqref{permica}
shows that close to extremality the permittivity remains negative, but there are certain regions where it can become
positive.

\subsubsection*{F1-Dp fluid}

The basic thermodynamic quantities are
\begin{equation}
\vep = \frac{\Omega_{(n+1)}}{16 \pi G} r_0^{n} ( 1 + n \cosh^2\alpha )
\end{equation}
\begin{equation}
\CT = \frac{n}{4\pi r_0 \cosh \alpha} \spa s = \frac{\Omega_{(n+1)}}{4G} r_0^{n+1} \cosh \alpha
\end{equation}
\begin{equation}
Q_{{\rm D}p} = \frac{\Omega_{(n+1)}}{16 \pi G} n r_0^{n} \cos \theta \cosh \alpha \sinh \alpha
\spa \Phi_{{\rm D}p} = \cos \theta \tanh \alpha
\end{equation}
\begin{equation}
\mc Q_{{\rm F}1} = \frac{\Omega_{(n+1)}}{16 \pi G} n r_0^{n} \sin \theta \cosh \alpha \sinh \alpha
\spa \Phi_{{\rm F}1} = \sin \theta \tanh \alpha
\end{equation}
Here  $n=7-p$. 

The permittivity of F1 charge density is
\beq
\label{permida}
\epsilon_\mathrm{F1}=\frac{1}{\sqrt{Q^2_\mathrm{Dp}+\QQ^2_\mathrm{F1}}} \tanh \alpha
\frac{(n-2) \frac{Q_\mathrm{Dp}^2}{Q_\mathrm{Dp}^2+\QQ^2_\mathrm{F1}}\sinh^2 \alpha-1}
{(n-2)\sinh^2\alpha-1}
~.
\eeq
At extremality the permittivity  reduces to 
\beq
\epsilon_\mathrm{F1}
=\frac{Q_{{\rm D}p}^2}{\lp Q_{{\rm D}p}^2+\mc
Q_\mathrm{F1}^2\rp^{3/2}}>0
\eeq
reflecting the fact that F1-D$p$ ($p \geq 1$) is a 1/2 BPS bound state. 
Thus for $p\geq 1$ the F1-D$p$ brane supports waves in which the F1 charge fluctuates in
directions transverse to the strings. Beyond the extremal limit the precise behavior of
$\epsilon_\mathrm{F1}$ depends on the value of $n$. For $n=1,2$ the F1 charge permittivity
is always positive. For $n>2$, $\epsilon_\mathrm{F1}>0$ except within an interval (see eq.~\eqref
{permidd}) where it is negative.

The extremal limit with a null wave along the F1 is obtained in
appendix~\ref{app:branesolns}. In terms of its parameters $q_{p,\alpha}$ and $\theta$ we have
\beq
\mc K=\frac{\Omega_{(n+1)}}{16 \pi G} n q_p\,,
\eeq
\beq
Q_{{\rm D}p} = \frac{\Omega_{(n+1)}}{16 \pi G} n
q_\alpha\cos\theta\,,\qquad
\Phi_{{\rm D}p} = \cos \theta,
\eeq
\beq
\mc Q_{{\rm F}1} = \frac{\Omega_{(n+1)}}{16 \pi G} n
q_\alpha\sin\theta, \qquad \Phi_{{\rm F}1} = \sin \theta\,.
\eeq

\subsection{Solutions for odd-spheres and products of odd-spheres \label{sec:multiodd}}

Since we have the explicit equations of state for several systems, we
can insert them into the generic solutions for products of odd-spheres
of sec.~\ref{subsec:oddq}. Blackfolds based on D0-D$p$ black branes with
spatially compact worldvolumes give new D0-brane-charged black holes
with D$p$-brane dipole, whereas blackfolds based on the F1-D$p$
solutions give dipole black holes with no conserved charge.

Our description of the solutions will be brief. We give the equilibrium
radii of the odd-spheres, and with these it is straightforward to
compute the thermodynamics of the various blackfold solutions using
\eqref{MJS}, \eqref{Phidef} for $M$, $J$, $S$ and $\Phi_H^{(p)}$ and
\eqref{Qq}, \eqref{PhiHq} for $\Phi_H^{(q)}$, $Q_q$. These values are
not much harder to obtain than those in appendix~\ref{app:STMBF} or in
refs.~\cite{Emparan:2009vd,Caldarelli:2010xz}, so we omit them.

Table \ref{oddspherelist} gives a summary of the types of horizons we
obtain in $D=10$ (with no KK torus).

\subsubsection{D0-D$p$}

For D0-D$p$ the radii $R_I$ of the wrapped spheres $S^{p_I}$ take the
equilibrium values \eqref{OmIRI},
\begin{equation}
\label{RD0Dp}
R_I = \sqrt{\frac{p_I}{p}}  \frac{\tanh \eta}{\Omega_I}  
\end{equation}
where the rapidity $\eta$ is given by 
\begin{equation}
\label{2chargerap1}
\sinh^2 \eta =  \frac{ p}{7-p} \frac{ 1 + (7-p) \sinh^2 \alpha_p }{1 +
\sinh^2 \alpha_0} \,.
\end{equation}
These solutions interpolate between D$p$ and smeared D0: for $\alpha_0
=0$ we recover the D$p$ blackfolds of Sec.~\ref{sec:oddsph} and for
$\alpha_p = 0$ we recover the D0-charged blackfolds of
\cite{Caldarelli:2010xz}, describing rotating charged black holes.

In the extremal limit the rapidity \eqref{2chargerap1} correctly reproduces \eqref{shetasolqp}.

\paragraph{Extremal D0-D6 black holes with finite area.}
The D0-D6 solutions have finite horizon area at extremality, and
therefore we obtain extremal, non-singular D0-charged black holes with
D6 dipole, with the variety of horizon topologies in
table~\ref{oddspherelist}. 

These are ten-dimensional black holes, but if the D6 wraps Kaluza-Klein
circles we find other extremal black holes in any dimension $4\leq D\leq
10$. All these black holes are rotating and non-supersymmetric, and as
we have seen, they are unstable to clumping the D0 charge. 

\begin{table}[t]
\begin{center}
\begin{tabular}{|c|c|c|c|} \hline 
 &  & Worldvolume  & $\perp$ Sphere \\ \hline \hline 
 & F1-D1 & (helical) $S^1$ & $s^7$ \\ \hline
 D0-D2 & F1-D2  & $\T^2$ & $s^6$ \\ \hline
 & F1-D3 & $S^3$\,,\quad $\T^3$ & $s^5$  \\ \hline
D0-D4 & F1-D4  &   $S^3 \times S^1$\,,\quad $\T^4$ & $s^4$ \\ \hline
& F1-D5 & $S^5$ \,,\quad $ S^3 \times \T^2$ & $s^3$  \\ \hline
D0-D6  & F1-D6 & $S^3  \times S^3 $ \,,\quad $S^5  \times S^1 $  & $s^2$ \\ \hline 
\hline
\end{tabular}
\bf\caption{\it A list of the allowed possibilities for blackfolds based
on D0-D$p$ and F1-D$p$ black branes wrapping products of
odd-spheres. The horizon topology of the corresponding black holes is
the product of the worldvolume and the transverse sphere $s^{n+1}$.
Eq.~\eqref{ellbound} limits the number $\ell$ of odd-spheres along the
worldvolume.
}
\label{oddspherelist}
\end{center}
\end{table}

\subsubsection{F1-D$p$}

For the case of F1-D$p$, with the fluid velocity along the F1, the
equilibrium solution is easily obtained from \eqref{1pextsol}. The total
rapidity is determined as
\beq
\sinh^2\eta=\frac{p}{n}+\lp\sin^2\theta +p\cos^2\theta\rp\sinh^2\alpha\,.
\eeq
This expression becomes simpler when $p=1$, which describes a F1-D1
bound state. The equilibrium condition for this is
\beq
\sinh^2\eta=\frac{1}{n}+\sinh^2\alpha
\eeq
which is of the same form as \eqref{soleta}, but now
$\sinh^2\alpha$
accounts for the tension of the bound state. Helical rings with these
charges are straightforward to obtain.

The extremal limit has a null wave parallel to the F1. The relations
derived in secs.~\ref{subsec:extq} and \ref{subsec:oddq} apply directly
to these configurations and they allow to easily compute the physical
parameters. We will see explicitly how this works in the configurations
that we study next.

\section{New extremal D1-D5-P black holes}\label{sec:D1D5P}

Black branes with D1-D5 charges are particularly prominent in string
theory. They are T-dual to the D0-D4 branes of the previous section, but
when their worldvolumes are curved to form a blackfold this duality
only applies locally, and is globally broken. Thus the black holes that
result are physically inequivalent\footnote{This is besides the fact
that the T-duality in supergravity smears the D0-D4 system along a
transverse direction. Smeared branes are unstable to
Gregory-Laflamme perturbations along the smearing direction \cite{Harmark:2005jk}.}.

Using our analysis in sec.~\ref{subsec:oddq} it is easy to construct
non-extremal D1-D5 blackfolds on products of odd-spheres, in particular
when the worldvolume velocity is aligned with the D1 current. Since
this velocity creates a local momentum on the worldvolume, the branes
are locally like a D1-D5-P system. Globally, they are black holes with
D1 and D5 dipoles and angular momentum. The main novelty that these
present is that their extremal limit, keeping the momentum finite (hence
null), has a regular horizon with finite area which is not obviously
unstable. Thus in the following we focus exclusively on extremal
solutions.

The solution for the extremal D1-D5-P system is given in
appendix~\ref{subsec:d1d5p}. This is a brane configuration with a null wave on it,
and in terms of the parameters $q_{1,5,p}$ of the solution we have
\begin{equation}
\vep = \frac{\pi}{4G}  ( q_1 + q_5 + q_p  )
\end{equation}
with charges and potentials
\begin{equation}
\mc Q_{1,5} = \frac{\pi}{4G}   q_{1,5}\,,\qquad \Phi_{1,5}=1\,,
\end{equation}
and momentum density
\beq
\mc K= \frac{\pi}{4G}   q_p\,.
\eeq
The stress tensor is
\beq
T_{ab} = {\mc K} l_a l_b -\sum_{q=1,5} \mc Q_q h^{(q)}_{ab}\,.
\eeq
The null vector in the geometry \eqref{d1d5pext} is $l = \partial_v$. 
The entropy density is finite
\begin{equation}
 s = \frac{\pi^2}{2 G}  \sqrt{ q_1 q_5 q_p }\,. 
\end{equation}

Applying our general construction of sec.~\ref{subsec:oddq} to this system, we
obtain 10D extremal rotating black holes with D1-D5 dipoles that have
regular horizons of topology $s^3\times S^5$ and $s^3\times \T^2\times
S^3$ --- eq.~\eqref{ellbound} forbids more than
four odd-sphere factors along this worldvolume. The general result \eqref{locK}
tells us that the condition of
equilibrium is
\beq
q_p=q_1+5q_5\,.
\eeq
The mass is
\beq
M=\frac{\pi}{2G}V_{(5)}(q_1+3q_5)
\eeq
where $V_{(5)}$ is the total volume that the blackfold wraps,
\beq
V_{(5)}=\left\{ \begin{array}{ll}
\pi^3 R^5& \quad\mathrm{for}~S^5\,,\\
8\pi^4 R_{S_1}R_{\tilde S^1} R_{S^3}^3&\quad\mathrm{for}~\T^2\times
S^3\,.
\end{array} \right.
\eeq
The angular momenta along the rotation directions of the spheres are
\beq	
J=\frac{\pi^4}{4G}R^6 q_p \qquad\mathrm{for}~S^5\,,
\eeq
\beq
J_I=\frac{\pi}{4G}V_{(5)}q_p R_I\sqrt{\frac{p_I}{5}} \qquad\mathrm{for}~\T^2\times S^3\,.
\eeq
The angular velocity on $S^5$ is $\Omega =R^{-1}$, and on each sphere of
$\T^2\times S^3$, $\Omega_I=R_I^{-1}\sqrt{p_I/5}$. 

On $S^5$ the dipole charges and potentials are
\beq
Q_{\rm D1}=\frac{\pi^3}{8G}R^4 q_1\,,\qquad Q_{\rm
D5}=\frac{\pi}{4G}q_5\,,\qquad
\Phi_H^{\rm D1}=2\pi R\,,\qquad \Phi_H^{\rm D5}=\pi^3 R^5
\eeq
and we can easily verify that
\beq
dM=\Omega dJ+\Phi_H^{\rm D1}dQ_{\rm D1}+\Phi_H^{\rm D5}dQ_{\rm D5}\,.
\eeq

If we wrap some of the directions of the five-brane on compact KK
circles, we obtain black holes in dimension $<10$. In $D=5$ we recover
the conventional static D1-D5-P black hole, but besides this one, all
these blackfolds are not supersymmetric. The D1 and the momentum (which
we take to be parallel) must wrap all the contractible cycles in the
non-compact part of the space in order to achieve equilibrium, but as
discussed in sec.~\ref{subsec:oddq}, the
KK circles are stabilized by topology and so it is not necessary that
the momentum and the D1 wrap them. Thus we get different
solutions depending on whether D1-P wrap the KK circles or not.

The new topologies that we can obtain with $10-D$ compact KK circles are
shown in table~\ref{table:d1d5p}. In $D=6$ we get black
rings,\footnote{For equal numbers of D1 and D5 branes, they are among the
solutions in sec.~\ref{sec:oddsph} with $N=2$, which have also been
constructed in \cite{Caldarelli:2010xz}.} which can have helical shape.

\begin{table}[t!]
\centering
\begin{tabular}{|c|c|c|}
\hline
Dimension (non-compact) & Worldvolume & $\perp$ Sphere   \\
\hline \hline
$D=10$  & $ S^5$\,,\quad   $S^3 \times \T^2  $ & $s^3$ \\ \hline
$D=9$   & $S^3\times S^1$\,,\quad  $\T^4$   & $s^3$  \\ \hline
$D=8$   & $ S^3$ \,,\quad $\T^3$ & $s^3$ \\ \hline
$D=7$   & $\T^2$ &  $s^3$      \\ \hline
$D=6$   & $S^1$ & $s^3$     \\ \hline
\end{tabular}
\bf\caption{\it A list of horizon topologies for stationary extremal
rotating black holes with D1-D5 dipoles in a spacetime with $D$
non-compact dimensions and $10-D$ compact KK circles.
We do not
distinguish whether the D1-P current wraps some of the
compact directions, which gives different kinds of black holes.}
\label{table:d1d5p}
\end{table}

The worldvolume of these extremal branes is marginally stable to
fluctuations of the D1 charge, and Gregory-Laflamme-type instabilities
seem to be absent. The brane being a tensile object, its purely
extrinsic perturbations are also stable. Therefore, there is no obvious
instability that afflicts these new extremal black holes. One
possible source of an instability might occur if the
coupling between intrinsic and extrinsic perturbations turned a marginal
mode into an unstable one. If this did not happen, these
could be the first stable, asymptotically flat, extremal,
non-supersymmetric black holes with non-spherical horizon topology in
$D\geq 6$ (in $D=5$ there are the black rings of \cite{Emparan:2004wy}).
They would also be the first asymptotically flat extremal
(supersymmetric or not) stable black holes with horizon topology other
than $S^{D-2}$ or $S^1\times S^2$. The six-dimensional black rings with
D1-D5 dipole appear as the most likely stable candidates for this
class.

\section{Discussion}\label{sec:discuss}

Tables \ref{II}, \ref{M}, \ref{oddspherelist} and \ref{table:d1d5p}
summarize the topologies of the horizons of the new black hole solutions
in string/M-theory that we have constructed. These tables are far from
being exhaustive lists of all the possibilities, even for a given kind
of brane. We have only studied the simplest class of solutions whose
worldvolume geometry is a product of round odd-spheres. However, these lists
are illustrative of the many new possibilities afforded by our
techniques.\footnote{See \cite{Compere:2010fm} for other recent work on
new charged and/or dipole black rings, and
on  near-horizon analyses of supersymmetric non-spherical black holes in string/M theory.}

Given that many of our solutions arise in string and M-theory, it is
natural to ask what their microscopic interpretation is. The answer
emerges directly from the nature of the blackfold construction. Since,
locally, the blackfold is well approximated by a flat black brane, it
follows that whatever the microscopic interpretation of the latter is,
it applies to the blackfold as well, only now with the microscopic brane
configuration being curved over a long length scale\footnote{This can correspond
to turning on a source for the corresponding CFT.}. This is
adequate to leading order in the blackfold
approximation. Corrections to the blackfold solution, from the extrinsic
curvature of the brane and the backreaction of the background, must be
matched with corrections in the microscopic picture, but in general
neither of these is easy to compute. This point of view was first
advocated and developed in \cite{Emparan:2004wy}, where five-dimensional black rings
were understood in terms of the microscopic interpretation of the black
string that they locally approximate, \eg\ the
Maldacena-Strominger-Witten CFT for the M5$\perp$M5$\perp$M5 system.

Thus, in the present case we can understand the microscopic entropy of
our new extremal black holes built by bending D1-D5-P or D0-D6 branes,
in terms of the microphysics of the latter as described in
\cite{Strominger:1996sh} and \cite{Emparan:2006it} respectively. There
is a caveat to this argument, though, which is most apparent in the
D1-D5-P system. In order to derive the $1+1$ CFT for the D1-D5 bound
state one assumes that the D5 directions transverse to the D1 wrap a
space much smaller than along the D1, so that
oscillations along the former are very massive and suppressed. In our
D1-D5-P blackfolds such a reduction occurs in $D=6$ (and perhaps also
when some odd-spheres are much smaller than a given $S^1$), but not in
general. In these other cases, even if the horizon entropy is nominally
accounted for in terms of the $1+1$ degrees of freedom, there is the
possibility that the system increase its entropy by exciting light
orthogonal modes. It would be interesting to study this for its possible
relevance to the dynamical stability problem. In light of this, the
six-dimensional rotating black rings (possibly helical) with D1-D5 dipoles are
probably the cleanest and best behaved of all the new black holes that we have
constructed.

Similar remarks apply to most other configurations, but it may be worth
commenting on the extremal singular solutions with the charge of a
D$p$-brane when they carry a null momentum wave. It may seem surprising
that a D-brane, whose worldvolume is locally Lorentz-invariant, can
support longitudinal waves on it. The resolution has been discussed for
the similar case of fundamental strings in \cite{Emparan:2004wy}: the
extremal null-wave branes are the macroscopic, coarse-grained
description of branes with an ensemble of travelling (transverse) waves
along their worldvolume. For a flat brane, the longitudinal momentum
wave can be resolved into transverse travelling waves, both in the
supergravity description and in the DBI or Nambu-Goto descriptions.
Ref.~\cite{BlancoPillado:2007iz} performs a detailed analysis of all
aspects of this picture, and its conclusions can be transported to our
present context. In particular, it is noted that when the worldvolume is
not flat but bent into a curved shape, the travelling waves will emit
gravitational radiation. In the coarse-grained version, the extremal
brane with a null wave is stationary at the classical level, but quantum
effects will make it decay slowly through the emission of superradiant
modes. Therefore, our new D1-D5-P extremal black holes, even if they are
classically stable, will decay at quantum-mechanical level. This is
indeed a generic feature of most extremal non-supersymmetric rotating
black holes \cite{Emparan:2007en,Dias:2007nj}.

\section*{Acknowledgements}
RE is grateful to Marco Caldarelli and David Mateos for useful discussions. NO thanks Andreas V. Pedersen for useful discussions. TH thanks Niels Bohr Institute for hospitality. TH, VN and NO thank the Galileo Galilei Institute for Theoretical Physics for the hospitality and the INFN for partial support during the completion of this work.
The work of RE is supported by MEC FPA2010-20807-C02-02, AGAUR 2009-SGR-168 and CPAN
CSD2007-00042 Consolider-Ingenio 2010. VN was partially supported by the
European Union grants FP7-REGPOT-2008-1-CreteHEPCosmo-228644 and
PERG07-GA-2010-268246. The work of NO is supported in part by the Danish National
Research Foundation project ``Black holes and their role in quantum gravity''.

\begin{appendix}

\section{Physical parameters for odd-sphere blackfold solutions\label{app:STMBF}}

For reference and as an illustration of how the explicit results are
obtained, here we collect the physical parameters for the blackfolds with
$p$-brane current on odd-spheres, with $p=2m+1$, as built in section \ref{sec:oddsph}.

We express the quantities in terms of the parameters $r_0$, $\alpha$,
and $R$, and introduce the shorthand notation $s\equiv\sinh\alpha$. 
Then,
\begin{subequations}
\begin{equation}
\label{tbrthermoo2}
M=\frac{V_{(p)} \Omega_{(n+1)}}{16 \pi G} \, r_0^{n}(n+p+1 + Nn(1+p)s^2),
\end{equation}
\beq
\label{SToddS} S=\frac{V_{(p)} \Omega_{(n+1)}}{4G}  r_0^{n+1}
\sqrt{\frac{(n+p+np s^2)(1+s^2)^N}{n}}\,,
\eeq
\beq
T = \frac{n}{4\pi} \sqrt{ \frac{n}{(n+p + N nps^2)(1+s^2)^N}}
\frac{1}{r_0}\,,
\eeq
\beq
\label{tbrthermo23}
J=\frac{V_{(p)} \Omega_{(n+1)}}{16\pi G}\,R \, 
r_0^{n}\sqrt{p(1+ N ns^2)(n+p + N np s^2)}\,,~~
\eeq
\beq
\Omega =  \sqrt{\frac{p(1+ N ns^2)}{n+p+ N nps^2} } \frac{1}{R}  
\, ,
\eeq
\begin{equation}
Q_p =  \frac{\Omega_{(n+1)}}{16 \pi G}  r_0^{n} \sqrt{N} n  s \sqrt{1+s^2}
\,,\qquad
\Phi^{(p)}_H = V_{(p)} \frac{\sqrt{N} s}{\sqrt{s^2+1}}
~.
\end{equation}
\end{subequations}
Here $V_{(p)} = R^p \Omega_{(p)}$ is the volume of a round $p$-sphere 
with radius $R$. $J$ and $\Omega$ are the angular momentum and velocity
along the vector $\partial/\partial\phi$ in \eqref{chiphi}. For this
solution $J$ is equally split
into components along each of the Cartan generators
$\partial/\partial\phi_i$ of the rotations of the sphere, \ie
$J_i=J/(m+1)$ and $\Omega_i=\Omega$.

As a check, these quantities agree with those derived in \cite{Emparan:2004wy} for the 5D dipole
ring at $n=p=1$, in the limit $\nu, \mu,\lambda  \rightarrow 0$ and $R \rightarrow \infty$, 
keeping fixed
\begin{equation}
r_0 = \nu R \spa r_0 \sinh^2 \alpha = \mu R \spa r_0(2 + N \sinh^2 \alpha) = \lambda R
\end{equation}
and redefining $ {\cal{Q}} \to 4 G Q_p$,
$\Phi \to \Phi^{(p)}_H/(4 G)$.

\section{Black brane solutions in type II string theory}
\label{app:branesolns}

As an aid to the reader, in this appendix we give the supergravity
solutions of the D0-D$p$, F1-D$p$ and extremal D1-D5-P brane
configurations in type II string theory. These are used in
Sec.~\ref{sec:twocharge} to extract the thermodynamic properties of the
corresponding effective fluids.

\subsubsection*{D0-D2 brane}

String frame metric (see \cite{Breckenridge:1996tt,Costa:1996zd}):
\begin{equation}
ds^2 = H ^{-\frac{1}{2}}   \left[ - f dt^2 +  D \sum_{i=1}^{2} dx_i^2  
+ H ( f^{-1} dr^2 + r^2 d\Omega_6^2 ) \right].
\end{equation}
Dilaton:
\begin{equation}
e^{2\phi} = D H^{\frac{1}{2}} \,.
\end{equation}
NSNS and RR potentials:
\begin{equation}
B_{12} =  \tan \theta [ D H^{-1} -1] 
\spa 
A_{0} = \coth \alpha  \sin \theta  [  H^{-1} - 1]  
\spa A_{012} = \cot \alpha  \sec \theta [ D  H^{-1} -1 ]\,.
\end{equation}
Functions:
\begin{equation}
f = 1 - \frac{r_0^5}{r^5} \spa H = 1 + \frac{r_0^5 \sinh^2 \alpha}{r^5} \spa 
D = ( H^{-1} \sin^2 \theta + \cos^2 \theta)^{-1} \,.
\end{equation}
\subsubsection*{D0-D4 brane}

String frame metric:
\begin{equation}
ds^2 = H_0^{-\frac{1}{2}}  H_4^{-\frac{1}{2}} \left[ - f dt^2 
+ H_0 \sum_{i=1}^{4} dx_i^2 + H_0 H_4 ( f^{-1} dr^2 + r^2 d\Omega_4^2 ) \right]\,.
\end{equation}
Dilaton:
\begin{equation}
e^{2\phi} = H_0^{\frac{3}{2}} H_4^{-\frac{1}{2}}\,.
\end{equation}
RR potentials:
\begin{equation}
A_{0} = \coth \alpha_0 [ H_0^{-1} - 1]  \spa A_{01234} = \coth \alpha_6 [  H_4^{-1} -1 ]\,.
\end{equation}
Functions:
\begin{equation}
f = 1 - \frac{r_0^3}{r^3} \spa H_i = 1 + \frac{r_0^3 \sinh^2 \alpha_i}{r^3}\,.
\end{equation}
for $i=0,4$.

\subsubsection*{D0-D6 brane}

Take the solution of the static dyonic KK black hole (first derived in
\cite{Dobiasch:1981vh,Chodos:1980df}, here we follow the presentation in
\cite{Larsen:1999pp}), uplift on 
$\T^6$ to M-theory and then reduce on the $y$-direction to type IIA string theory.
In order to parallel the notation above, we also redefine $2m= r_0$, $q= r_0 \cosh^2 \alpha_0$
and $p= r_0 \cosh^2 \alpha_6$. This means 
\begin{equation}
2 Q = r_0 \cosh \alpha_0 \sinh \alpha_0 \sqrt{ \frac{ \cosh^2 \alpha_0 +1}{\cosh^2 \alpha_0 
+ \cosh^2 \alpha_6} } \equiv q_0
\end{equation}
\begin{equation}
2 P = r_0 \cosh \alpha_6 \sinh \alpha_6 \sqrt{ \frac{ \cosh^2 \alpha_6 +1}{\cosh^2 \alpha_0 
+ \cosh^2 \alpha_6} }\equiv q_6
\end{equation}

\noindent
String frame metric:
\begin{equation}
ds^2 = H_0^{-\frac{1}{2}}  H_6^{-\frac{1}{2}} \left[ - f dt^2 + H_0 \sum_{i=1}^{6} dx_i^2 
+ H_0 H_6 ( f^{-1} dr^2 + r^2 d\Omega_2^2 ) \right]\,.
\end{equation}
Dilaton:
\begin{equation}
e^{2\phi} = H_0^{\frac{3}{2}} H_6^{-\frac{3}{2}}\,.
\end{equation}
RR potentials:
\begin{equation}
A_{0} =  - \frac{q_0}{r} \left[ 1 + \frac{r_0}{2r} \sinh^2 \alpha_6 \right] H_0^{-1}
\spa A_{\phi} = - q_6 \cos \theta\,.
\end{equation}
Functions:
\begin{equation}
f = 1 - \frac{r_0}{r} \spa H_i = 1 + \frac{r_0\sinh^2 \alpha_i}{r} + \frac{r_0^2\cosh^2 \alpha_i}{2 r^2}
 \frac{ \sinh^2 \alpha_0 \sinh^2 \alpha_6}{ \cosh^2 \alpha_0 + \cosh^2 \alpha_6}\,.
\end{equation}
for $i=0,6$.

One may compute
\begin{equation}
F_{0r} = \frac{q_0}{r^2} \frac{H_6}{H_0^2} \,,\qquad F_{\theta \phi} = q_6 \sin \theta
\end{equation}
We use 
$ \tilde F_{\mu_0 \ldots \mu_p r } = -
\sqrt{-g} e^{a \phi}
\epsilon_{\nu_{p+2} \ldots \nu_{D-1} \mu_0 \ldots \mu_p r  } F^{ \nu_{p+2} \ldots \nu_{D-1} }$ 
to compute the dual of the magnetic field strength (in Einstein frame) with $a=3/2$ for the case at hand. This gives
\begin{equation}
\tilde F_{01 \ldots 6 r} = \frac{q_6}{r^2} \frac{H_0}{H_6^2}
\end{equation}
and hence
\begin{equation}
A_{01 \ldots 6} =  - \frac{q_6}{r} \left[ 1 + \frac{r_0}{2r} \sinh^2 \alpha_0 \right] H_6^{-1}
\end{equation}

Note also that for one of the  $\alpha_i =0$, the results agree with the (smeared on $\T^6$) 
D0-brane or the D6-brane respectively.

\subsubsection*{F1-D$p$ brane}

We have $n=7-p$. The construction of F1-D$p$ branes was first
explained in \cite{Costa:1996zd}. The
explicit non-extremal solution that we use here was
given in \cite{Harmark:2000wv}.

\vspace{0.3cm}
\noindent
String frame metric:
\begin{equation}
ds^2 = D^{-\frac{1}{2}} H^{-\frac{1}{2}} \left[ - f dt^2 + dx_1^2 \right] + D^{\frac{1}{2}} H^{-\frac{1}{2}} 
\sum_{i=2}^p dx_i^2 + D^{-\frac{1}{2}} H^{\frac{1}{2}} \left[  f^{-1} dr^2 + 
r^2 d\Omega_{n+1}^2 \right]\,.
\end{equation}
Dilaton:
\begin{equation}
e^{2\phi} = D^{\frac{p-5}{2}} H^{\frac{3-p}{2}}\,.
\end{equation}
NSNS and RR potentials:
\begin{equation}
B_{01} = \sin \theta ( H^{-1} -1 ) \coth \alpha
\spa A_{2\cdots p} = \tan \theta ( H^{-1} D -1 )
\spa A_{01 \cdots p }= \cos \theta D (H^{-1} -1) \coth \alpha\,.
\end{equation}
Functions:
\begin{equation}
f = 1 - \frac{r_0^n}{r^n} \spa H = 1 + \frac{r_0^n \sinh^2 \alpha}{r^n} \spa D^{-1} 
= \cos^2 \theta + \sin^2 \theta H^{-1}\,.
\end{equation}

In the extremal limit we send $r_0\to 0$ and $\alpha\to\infty$ keeping finite
\beq
q_\alpha=r_0^n\sinh^2\alpha\,.
\eeq
If at the same time, we boost the system along $x_1$ with rapidity
$\eta\to\infty$ and keep finite
\beq
q_p=r_0^n\sinh^2\eta
\eeq
we obtain (in string frame)
\beq
ds^2 = D^{-\frac{1}{2}} H^{-\frac{1}{2}} \left[ - du
dv+\frac{q_p}{r^n}dv^2 \right] + D^{\frac{1}{2}} H^{-\frac{1}{2}} 
\sum_{i=2}^p dx_i^2 + D^{-\frac{1}{2}} H^{\frac{1}{2}} \left[  dr^2 + 
r^2 d\Omega_{n+1}^2 \right]\,.
\eeq
with $u=t+x_1$ and $v= t-x_1$ and
\begin{equation}
H = 1 + \frac{q_\alpha}{r^n}\,, \qquad D^{-1} 
= \cos^2 \theta + \sin^2 \theta H^{-1}\,.
\end{equation}

\subsubsection*{Extremal D1-D5-P brane}\label{subsec:d1d5p}

\vspace{0.3cm}
\noindent
String frame metric:
\begin{equation}\label{d1d5pext}
ds^2 =  H_5^{-\frac{1}{2}} H_1^{-\frac{1}{2}}
\left[ - du dv + \frac{q_p}{r^2} dv^2  +  H_1 \sum_{i=1}^4 dx_i^2 +
H_5 H_1 ( dr^2 + r^2 d\Omega_3^2 ) \right]\,.
\end{equation}
where
\begin{equation}
H_1 = 1 + \frac{q_1}{r^2} \spa H_5 = 1 + \frac{q_5}{r^2} 
\end{equation}
and the lightcone coordinates are $u=t+z$ and $v= t-z$.

Dilaton:
\begin{equation}
e^{2\phi} = H_5^{-1} H_1\,.
\end{equation}
RR potentials:
\begin{equation}
 A_{01234z} =  H_5^{-1} -1 \spa 
A_{0z} =   H_1^{-1} -1\,.
\end{equation}

\section{Details on isothermal permittivities \label{app:iso}} 

We collect some of the details of the isothermal permittivities computed for the effective
D$p$-brane fluids with $q$-brane charge that were used in Section~\ref{sec:2chargefluid}.
Similar quantities were computed in \cite{Friess:2005tz}.

\subsubsection*{D0-D4 fluid}
For the D0-D4 fluid the natural non-extremality parameter is 
\beq
\label{permibb}
x \equiv 1-\left(\frac{\QQ_\mathrm{D0}+Q_\mathrm{D4}}{\varepsilon}\right)^2
\eeq
which can be expressed in terms of $\alpha_0$ as 
\beq
\label{permibc}
x=1-\frac{9}{4} \frac{\left( 1+\frac{Q_\mathrm{D4}}{\QQ_\mathrm{D_0}} \right)^2
\left(\cosh^2(2\alpha_0)-1\right)}
{\left(4-3\frac{Q^2_\mathrm{D_4}}{\QQ^2_\mathrm{D_0}}+3 \cosh(2\alpha_0)
+3 \frac{Q^2_\mathrm{D_4}}{\QQ^2_\mathrm{D_0}} \cosh^2(2\alpha_0)
\right)^2}
~.
\eeq
Also  in terms of $\alpha_0$ the permittivity of D0 charge \eqref{permiba} takes the form 
\beq
\label{permibd}
\epsilon_\mathrm{D0}=\frac{1}{\QQ_\mathrm{D0}}
\frac{\tanh \alpha_0\left( -3+\sqrt{1+  \frac{Q^2_\mathrm{D_4}}{\QQ^2_\mathrm{D_0}}
\sinh^2(2\alpha_0)}\right)}
{2(1-\sinh^2\alpha_0)-(1+8 \sinh^2\alpha_0)\left(-1
+\sqrt{1+ \frac{Q^2_\mathrm{D_4}}{\QQ^2_\mathrm{D_0}}\sinh^2(2\alpha_0)}\right)}
~.
\eeq
When we vary $\alpha_0$ (and therefore $x$) at fixed $\QQ_\mathrm{D_0},
Q_\mathrm{D4}$, we find that $\epsilon_\mathrm{D0}<0$ except in a finite range
$\alpha_0\in (\alpha_-,\alpha_+)$ where $\epsilon_\mathrm{D0}>0$. $\alpha_-$ is the value
of $\alpha_0$ where the denominator of the expression \eqref{permibd} vanishes and
\beq
\label{permide}
\sinh^2(2\alpha_+)=8\left(\frac{\QQ_\mathrm{D0}}{Q_\mathrm{D4}}\right)^2
~.
\eeq

\subsubsection*{D0-D6 fluid}
For the D0-D6 fluid  the D0 charge permittivity n terms of $\alpha_0, \alpha_6$ is given by 
\beq
\label{permica}
\epsilon_\mathrm{D0}=\frac{1}{\QQ_\mathrm{D0}}
\frac{\sinh\alpha_0 \cosh^2\alpha_6 (\cosh^2\alpha_0+\cosh^2\alpha_6)}
{(1+\cosh^2 \alpha_6)\sqrt{(1+\cosh^2\alpha_0)(\cosh^2\alpha_0+\cosh^2\alpha_6)}}
\frac{\NN(\alpha_0,\alpha_6)}{\DD(\alpha_0,\alpha_6)}
\eeq
where
\bea
\label{permicb}
\NN(\alpha_0,\alpha_6)&=&\cosh(3\alpha_0)\left( 72-9\cosh(2\alpha_6)+\cosh(6\alpha_6)\right)
\nonumber\\
&&+\cosh\alpha_0 \left( 412+263 \cosh(2\alpha_6)+28 \cosh(4\alpha_6)+\cosh(6\alpha_6)\right)
\nonumber\\
&&-16 {\rm sech} \alpha_0 \cosh^2\alpha_6 (-5+\cosh(2\alpha_6))(3+\cosh(2\alpha_6))
~,
\eea
\bea
\label{permicc}
\DD(\alpha_0,\alpha_6)&=&71+112 \cosh(2\alpha_0)+37 \cosh(4\alpha_0)+4 \cosh(6\alpha_0)
\nonumber\\
&&+\cosh(2\alpha_6)\left( 99+129 \cosh(2\alpha_0)+27 \cosh(4\alpha_0)+\cosh(6\alpha_0) \right)
\nonumber\\
&&-\cosh(4\alpha_6) \left( -27-13 \cosh(2\alpha_0)+7 \cosh(4\alpha_0)+\cosh(6\alpha_0) \right)
\nonumber\\
&&-4\sinh^2 \alpha_0 (2+\cosh(2\alpha_0))\cosh(6\alpha_6)
~.
\eea

\subsubsection*{F1-Dp fluid}
For the F1-D$p$ fluid the natural non-extremality parameter is 
(in analogy to  \eqref{permiab} for the D0-D2 system),
\beq
\label{permidb}
x \equiv 1-\frac{\QQ_\mathrm{F1}^2+Q^2_\mathrm{Dp}}{\varepsilon^2}
\eeq
so that
\beq
\label{permidc}
\cosh^2\alpha=\frac{n+2-2x+\sqrt{(n+2)^2-4(n+1) x}}{2nx} ~.
\eeq
The permittivity of F1 charge density \eqref{permida} exhibits different behavior depending 
on the value of $n$. For $n=1,2$ the F1 charge permittivity is always positive. For $n>2$, 
$\epsilon_\mathrm{F1}>0$ except within the interval
\beq
\label{permidd}
x\in \left (
\frac{1+n(n+2) \left(1+\frac{1}{n-2}
\frac{\QQ_\mathrm{F1}^2+Q^2_\mathrm{Dp}} {Q_\mathrm{Dp}^2} \right)}
{\left(1+n\left(1+\frac{1}{n-2}
\frac{\QQ_\mathrm{F1}^2+Q^2_\mathrm{Dp}} {Q_\mathrm{Dp}^2} \right) \right)^2} ,
\frac{(n-2)(n-2+n(n-1)(n+2)}{(n^2-2)^2}
\right)
\eeq
where it is negative.

\end{appendix}

\addcontentsline{toc}{section}{References}



\begin{thebibliography}{99}

\bibitem{Emparan:2009cs}
  R.~Emparan, T.~Harmark, V.~Niarchos, N.~A.~Obers,
  ``World-Volume Effective Theory for Higher-Dimensional Black Holes,''
  Phys.\ Rev.\ Lett.\  {\bf 102 } (2009)  191301.
  [arXiv:0902.0427 [hep-th]].

\bibitem{Emparan:2009at}
  R.~Emparan, T.~Harmark, V.~Niarchos, N.~A.~Obers,
  ``Essentials of Blackfold Dynamics,''
  JHEP {\bf 1003} (2010)  063.
  [arXiv:0910.1601 [hep-th]].

\bibitem{Caldarelli:2010xz}
  M.~M.~Caldarelli, R.~Emparan, B.~Van Pol,
  ``Higher-dimensional Rotating Charged Black Holes,''
  JHEP {\bf 1104} (2011)  013.
  [arXiv:1012.4517 [hep-th]].

\bibitem{Grignani:2010xm}
  G.~Grignani, T.~Harmark, A.~Marini, N.~A.~Obers, M.~Orselli,
  ``Heating up the BIon,'' JHEP {\bf 1106} (2011) 058.
  [arXiv:1012.1494 [hep-th]];
  ``Thermodynamics of the hot BIon,''
To appear in Nucl.~Phys.~B.  [arXiv:1101.1297 [hep-th]].

\bibitem{Carter:2000wv}
  B.~Carter,
  ``Essentials of classical brane dynamics,''
  Int.\ J.\ Theor.\ Phys.\  {\bf 40 } (2001)  2099-2130.
  [gr-qc/0012036].


\bibitem{Camps:2008hb}
  J.~Camps, R.~Emparan, P.~Figueras, S.~Giusto, A.~Saxena,
  ``Black Rings in Taub-NUT and D0-D6 interactions,''
  JHEP {\bf 0902 } (2009)  021.
  [arXiv:0811.2088 [hep-th]].


\bibitem{Caldarelli:2008pz}
  M.~M.~Caldarelli, R.~Emparan, M.~J.~Rodr{\'\i}guez,
  ``Black Rings in (Anti)-deSitter space,''
  JHEP {\bf 0811} (2008) 011.
  [arXiv:0806.1954 [hep-th]].

  J.~Armas, N.~A.~Obers,
  ``Blackfolds in (Anti)-de Sitter Backgrounds,''
  Phys.\ Rev.\  {\bf D83} (2011) 084039.
  [arXiv:1012.5081 [hep-th]].


\bibitem{Emparan:2004wy}
  R.~Emparan,
  ``Rotating circular strings, and infinite nonuniqueness of black rings,''
  JHEP {\bf 0403 } (2004)  064.
  [hep-th/0402149].


\bibitem{Copsey:2005se}
  K.~Copsey, G.~T.~Horowitz,
  ``The Role of dipole charges in black hole thermodynamics,''
  Phys.\ Rev.\  {\bf D73}, 024015 (2006).
  [hep-th/0505278].

\bibitem{Emparan:2009vd}
  R.~Emparan, T.~Harmark, V.~Niarchos, N.~A.~Obers,
  ``New Horizons for Black Holes and Branes,''
  JHEP {\bf 1004} (2010)  046.
  [arXiv:0912.2352 [hep-th]].
  
\bibitem{Gubser:2000nd}
  S.~S.~Gubser,
  ``Curvature singularities: The Good, the bad, and the naked,''
  Adv.\ Theor.\ Math.\ Phys.\  {\bf 4 } (2000)  679-745.
  [hep-th/0002160].



\bibitem{Emparan:2010ni}
  R.~Emparan, S.~Ohashi, T.~Shiromizu,
  ``No-dipole-hair theorem for higher-dimensional static black holes,''
  Phys.\ Rev.\  {\bf D82 } (2010)  084032.
  [arXiv:1007.3847 [hep-th]].

\bibitem{Horowitz:2004je}
  G.~T.~Horowitz, H.~S.~Reall,
  ``How hairy can a black ring be?,''
  Class.\ Quant.\ Grav.\  {\bf 22 } (2005)  1289-1302.
  [hep-th/0411268].

\bibitem{BlancoPillado:2007iz}
  J.~J.~Blanco-Pillado, R.~Emparan, A.~Iglesias,
  ``Fundamental Plasmid Strings and Black Rings,''
  JHEP {\bf 0801 } (2008)  014.
  [arXiv:0712.0611 [hep-th]].

\bibitem{Emparan:2008qn}
  R.~Emparan,
  ``Exact Microscopic Entropy of Non-Supersymmetric Extremal Black Rings,''
  Class.\ Quant.\ Grav.\  {\bf 25 } (2008)  175005.
  [arXiv:0803.1801 [hep-th]].

\bibitem{Elvang:2004xi}
  H.~Elvang, R.~Emparan, P.~Figueras,
  ``Non-supersymmetric black rings as thermally excited supertubes,''
  JHEP {\bf 0502 } (2005)  031.
  [hep-th/0412130].


\bibitem{Gregory:1994bj}
  R.~Gregory, R.~Laflamme,
  ``The Instability of charged black strings and p-branes,''
  Nucl.\ Phys.\  {\bf B428 } (1994)  399-434.
  [hep-th/9404071].

  T.~Harmark, V.~Niarchos, N.~A.~Obers,
  ``Instabilities of black strings and branes,''
  Class.\ Quant.\ Grav.\  {\bf 24} (2007) R1-R90.
  [hep-th/0701022].


\bibitem{Camps:2010br}
  J.~Camps, R.~Emparan, N.~Haddad,
  ``Black Brane Viscosity and the Gregory-Laflamme Instability,''
  JHEP {\bf 1005 } (2010)  042.
  [arXiv:1003.3636 [hep-th]].

\bibitem{Gregory:1994tw}
  R.~Gregory, R.~Laflamme,
  ``Evidence for stability of extremal black p-branes,''
  Phys.\ Rev.\  {\bf D51 } (1995)  305-309.
  [hep-th/9410050].

\bibitem{Reall:2001ag}
  H.~S.~Reall,
  ``Classical and thermodynamic stability of black branes,''
  Phys.\ Rev.\  {\bf D64 } (2001)  044005.
  [hep-th/0104071].


\bibitem{Hirayama:2002hn}
  T.~Hirayama, G.~Kang, Y.~Lee,
  ``Classical stability of charged black branes and the Gubser-Mitra conjecture,''
  Phys.\ Rev.\  {\bf D67 } (2003)  024007.
  [hep-th/0209181].

\bibitem{Gubser:2002yi}
  S.~S.~Gubser, A.~Ozakin,
  ``Universality classes for horizon instabilities,''
  JHEP {\bf 0305 } (2003)  010.
  [hep-th/0301002].

\bibitem{Kang:2004hm}
  G.~Kang, J.~Lee,
  ``Classical stability of black D3 branes,''
  JHEP {\bf 0403 } (2004)  039.
  [hep-th/0401225].


\bibitem{Miyamoto:2006nd}
  U.~Miyamoto, H.~Kudoh,
  ``New stable phase of non-uniform charged black strings,''
  JHEP {\bf 0612 } (2006)  048.
  [gr-qc/0609046].

\bibitem{Miyamoto:2007mh}
  U.~Miyamoto,
  ``Analytic evidence for the Gubser-Mitra conjecture,''
  Phys.\ Lett.\  {\bf B659 } (2008)  380-384.
  [arXiv:0709.1028 [hep-th]].


\bibitem{Gubser:2000mm}
  S.~S.~Gubser, I.~Mitra,
  ``The Evolution of unstable black holes in anti-de Sitter space,''
  JHEP {\bf 0108} (2001) 018.
  [hep-th/0011127].


\bibitem{Buchel:2007mf}
  A.~Buchel,
  ``Bulk viscosity of gauge theory plasma at strong coupling,''
  Phys.\ Lett.\  {\bf B663 } (2008)  286-289.
  [arXiv:0708.3459 [hep-th]].

\bibitem{Kanitscheider:2009as}
  I.~Kanitscheider, K.~Skenderis,
  ``Universal hydrodynamics of non-conformal branes,''
  JHEP {\bf 0904 } (2009)  062.
  [arXiv:0901.1487 [hep-th]].


\bibitem{Mateos:2011tv}
  D.~Mateos, D.~Trancanelli,
  ``Thermodynamics and Instabilities of a Strongly Coupled Anisotropic Plasma,''
  [arXiv:1106.1637 [hep-th]];
  ``The anisotropic N=4 super Yang-Mills plasma and its instabilities,''
  [arXiv:1105.3472 [hep-th]].

\bibitem{Harmark:2005jk}
  T.~Harmark, V.~Niarchos, N.~A.~Obers,
  ``Instabilities of near-extremal smeared branes and the correlated stability conjecture,''
  JHEP {\bf 0510} (2005) 045 .
  [hep-th/0509011].



\bibitem{Compere:2010fm}
  G.~Compere, S.~de Buyl, S.~Stotyn, A.~Virmani,
  ``A General Black String and its Microscopics,''
  JHEP {\bf 1011} (2010) 133.
  [arXiv:1006.5464 [hep-th]].
 %

  B.~Kleihaus, J.~Kunz, K.~Schnulle,
  ``Charged Balanced Black Rings in Five Dimensions,''
  Phys.\ Lett.\  {\bf B699} (2011) 192-198.
  [arXiv:1012.5044 [hep-th]].
%

  I.~Bena, S.~Giusto, C.~Ruef,
  ``A Black Ring with two Angular Momenta in Taub-NUT,''
    [arXiv:1104.0016 [hep-th]].

  H.~K.~Kunduri, J.~Lucietti,
  ``Static near-horizon geometries in five dimensions,''
  Class.\ Quant.\ Grav.\  {\bf 26 } (2009)  245010.
  [arXiv:0907.0410 [hep-th]];
  ``Constructing near-horizon geometries in supergravities with hidden symmetry,''
  [arXiv:1104.2260 [hep-th]].

  J.~Gutowski, G.~Papadopoulos,
  ``Heterotic Black Horizons,''
  JHEP {\bf 1007 } (2010)  011.
  [arXiv:0912.3472 [hep-th]];
  ``Static M-horizons,''
    [arXiv:1106.3085 [hep-th]].



\bibitem{Strominger:1996sh}
  A.~Strominger, C.~Vafa,
  ``Microscopic origin of the Bekenstein-Hawking entropy,''
  Phys.\ Lett.\  {\bf B379 } (1996)  99-104.
  [hep-th/9601029].

\bibitem{Emparan:2006it}
  R.~Emparan, G.~T.~Horowitz,
  ``Microstates of a Neutral Black Hole in M Theory,''
  Phys.\ Rev.\ Lett.\  {\bf 97 } (2006)  141601.
  [hep-th/0607023].


\bibitem{Emparan:2007en}
  R.~Emparan, A.~Maccarrone,
  ``Statistical description of rotating Kaluza-Klein black holes,''
  Phys.\ Rev.\  {\bf D75 } (2007)  084006.
  [hep-th/0701150].

\bibitem{Dias:2007nj}
  O.~J.~C.~Dias, R.~Emparan, A.~Maccarrone,
  ``Microscopic theory of black hole superradiance,''
  Phys.\ Rev.\  {\bf D77 } (2008)  064018.
  [arXiv:0712.0791 [hep-th]].


\bibitem{Breckenridge:1996tt}
  J.~C.~Breckenridge, G.~Michaud, R.~C.~Myers,
  ``More D-brane bound states,''
  Phys.\ Rev.\  {\bf D55 } (1997)  6438-6446.
  [hep-th/9611174].

\bibitem{Costa:1996zd}
  M.~S.~Costa, G.~Papadopoulos,
  ``Superstring dualities and p-brane bound states,''
  Nucl.\ Phys.\  {\bf B510 } (1998)  217-231.
  [hep-th/9612204].


\bibitem{Dobiasch:1981vh}
  P.~Dobiasch, D.~Maison,
  ``Stationary, Spherically Symmetric Solutions Of Jordan's Unified Theory Of Gravity And Electromagnetism,''
  Gen.\ Rel.\ Grav.\  {\bf 14 } (1982)  231-242.

\bibitem{Chodos:1980df}
  A.~Chodos, S.~L.~Detweiler,
  ``Spherically Symmetric Solutions In Five-dimensional General Relativity,''
  Gen.\ Rel.\ Grav.\  {\bf 14 } (1982)  879.
  

\bibitem{Larsen:1999pp}
  F.~Larsen,
  ``Rotating Kaluza-Klein black holes,''
  Nucl.\ Phys.\  {\bf B575 } (2000)  211-230.
  [hep-th/9909102].

\bibitem{Harmark:2000wv}
  T.~Harmark,
  ``Supergravity and space-time noncommutative open string theory,''
  JHEP {\bf 0007 } (2000)  043.
  [hep-th/0006023].

\bibitem{Friess:2005tz}
  J.~J.~Friess, S.~S.~Gubser,
  ``Instabilities of D-brane bound states and their related theories,''
  JHEP {\bf 0511 } (2005)  040.
  [hep-th/0503193].



  

    
\end{thebibliography}
\end{document}